\documentclass[12pt,letterpaper,preprint]{aastex}

\usepackage{amsmath}
\usepackage{amsfonts}
\usepackage{amssymb}
\usepackage{float}
\usepackage{graphicx}
\usepackage[toc,page]{appendix}
\usepackage[section]{placeins}
\usepackage{morefloats}
\usepackage{mathrsfs}
\usepackage{amsfonts}

\def\harm{{\tt harm}}
\def\grm{{\tt grmonty}}
\def\bhl{{\tt bhlight}}

\def\munit{{\mathcal{M}}}
\def\tunit{{\mathcal{T}}}
\def\lunit{{\mathcal{L}}}

\def\order{\mathcal{O}}
\def\gdet{\sqrt{-g}}
\def\d{\mathrm{d}}
\def\D{\mathrm{D}}
\def\<{\langle}
\def\>{\rangle}
\def\Ledd{L_{\mathrm{Edd}}}

\begin{document}

\title{\bhl: General Relativistic Radiation Magnetohydrodynamics with Monte Carlo Transport}
\author{B. R. Ryan}
\affil{Department of Astronomy, University of Illinois, 1110 West Green Street, Urbana, IL, 61801}
\author{J. C. Dolence}
\affil{Los Alamos National Laboratory, Los Alamos, NM 87545, USA}
\author{C. F. Gammie}
\affil{Department of Astronomy, University of Illinois, 1110 West Green Street, Urbana, IL, 61801}
\affil{Department of Physics, University of Illinois, 1110 West Green Street, Urbana, IL, 61801}

\begin{abstract}

We present \bhl, a numerical scheme for solving
the equations of general relativistic radiation
magnetohydrodynamics (GRRMHD) using a direct
Monte Carlo solution of the frequency-dependent radiative transport
equation. \bhl~is designed to evolve black hole accretion flows at
intermediate accretion rate, in the regime between the classical
radiatively efficient disk and the radiatively inefficient
accretion flow (RIAF), in which global radiative effects play a sub-dominant but
non-negligible role in disk dynamics. We describe the governing
equations, numerical method, idiosyncrasies of our
implementation, and a suite of test and convergence results.  We also
describe example applications to radiative Bondi accretion and to a
slowly accreting Kerr black hole in axisymmetry.

\end{abstract}

\section{Introduction}

Many of the brightest objects in the universe, including quasars and the lesser active galactic nuclei, stellar-mass black hole binaries, and gamma-ray bursts, are likely the results of black hole accretion driven at least in part by the magnetorotational instability (MRI; \citealt{mri}). The structure of the luminous plasma surrounding the black hole remains uncertain (see the recent review of \citealt{begelman2014}), because it is difficult to resolve and because of physical complexity: relativistic gravity, turbulence in a magnetized plasma, and radiation transport all play some role in determining accretion flow structure.

Nonetheless, accreting black holes may be partially classified according to the ratio of their luminosity $L$ to the Eddington luminosity, $\Ledd \equiv 4 \pi G M c/\kappa_{es}$.  

For $L \gtrsim 10^{-2}\Ledd$, radiation is dynamically important.  Up to $L \sim \Ledd$, this regime can be modeled by the aligned thin $\alpha$ disk model of \cite{shakura1973}, in which the disk is geometrically thin and optically thick, and in which radiation pressure exceeds gas pressure at radii where most of the disk luminosity is produced. The radiative efficiency of the accretion flow, $\eta(a_*) \equiv L/(\dot{M} c^2)$, is expected to be approximately constant and determined by the dimensionless black hole spin $-1 < a_* < 1$.  It is common to describe the accretion rate $\dot{M}$ in units of an Eddington rate defined using a nominal efficiency $\eta = 0.1$: $\dot{m} = \eta\dot{M} c^2/\Ledd$. For $L \gtrsim\Ledd$, the flow is expected to resemble the slim disk solution of \cite{abramowicz1988}, in which the flow becomes geometrically thick as a result of long radiation diffusion times.  An obstacle to fully modeling the innermost, relativistic regions of flows with $\dot{m} \gtrsim 10^{-2}$ is the need for an efficient relativistic radiation hydrodynamics scheme that can operate in both the optically thick (disk midplane) and optically thin (disk atmosphere, corona, funnel) regimes.  

For $L \ll \Ledd$, or $\dot{m} \ll 1$, accretion is likely to occur through a radiatively inefficient accretion flow (RIAF or ADAF; see the recent review by Yuan \& Narayan 2014), in which the cooling time of a parcel of plasma is much longer than the time required for it to fall into the black hole.  Radiation plays no role in determining the flow structure.  RIAFs are believed to be geometrically thick, optically thin, collisionless plasmas that are at least partially supported by rotation.  RIAFs are commonly modeled numerically using relativistic magnetohydrodynamic (MHD) codes, but it is unclear how well the fluid model describes the dynamics of the magnetized, collisionless plasma.  It is also unclear how best to model the electrons, which are collisionally decoupled from the ions and determine the radiative properties of the plasma. However, local models, particularly numerical kinetic calculations, are beginning to constrain the electron distribution function in this regime (e.g. \citealt{kunz2014}, \citealt{riquelme2014}, \citealt{sironi2014}).

Between thin disks and RIAFs lies an intermediate regime in which radiation plays a modest role in the accretion flow; this configuration may be thought of as a RIAF perturbed by radiative effects. ADAF solutions evaluated at these accretion rates indicate a flow that is optically thin to Compton scattering ($\tau \sim 10^{-5}-10^{-3}$; \citealt{yuan2006}), and optically thick only to synchrotron self-absorption at long wavelengths.  As accretion rate increases the first non-negligible radiation-plasma interactions are expected to be Compton cooling and synchrotron cooling.  For example, M87's central black hole, an object of interest for the Event Horizon Telescope (\citealt{doeleman2009}),  is expected to reside in this intermediate regime ($\dot{m} \lesssim 6.3\times10^{-6}$, based on a RIAF model; \citealt{kuo2014}), in \cite{moscibrodzka2011} and \cite{dexter2012}. Such systems exhibit nonlinear evolution of coupled gas and radiation in strong gravity; predictive modeling is our primary motivation for \bhl, a numerical scheme for general relativistic radiation magnetohydrodynamics. 

In the nonrelativistic and $\mathcal{O}(v/c)$ regimes, many numerical methods have been developed to solve the radiation hydrodynamics equations (see the comprehensive review of  \citealt{castor2004}), including flux-limited diffusion.  Of particular relevance to black hole accretion flows is recent work on accretion in the near-Eddington regime using flux-limited diffusion (\citealt{hirose2009,hirose2014}) and using the more accurate short characteristics method, in which specific intensity is discretized in angle for each grid zone (\citealt{stone1992, jiang2012, jiang2014a, jiang2014b}) and one obtains a full solution to the grey transfer equation.

Close to the event horizon special and general relativistic effects can produce order unity variations in the intensity.  These effects are particularly important for rapidly rotating black holes.  Numerical schemes for solving the equations of general relativistic radiation MHD (GRRMHD) have only been developed in the last few years.  All are frequency-integrated and use approximate closure schemes, including the Eddington approximation (\citealt{farris2008}, \citealt{zanotti2011}, and \citealt{fragile2012}) and low-order truncated moment closure (\citealt{shibata2011}, \citealt{sadowski2013}, and \citealt{mckinney2013}).  These schemes are formally accurate at high optical depth, but not for general flows at the modest optical depths relevant to black holes in the intermediate accretion rate regime.

An alternative treatment of radiation, the Monte Carlo technique, has long been used for solving the full frequency-dependent transport equation without recourse to any closure model. Several radiation hydrodynamics schemes have recently been employed in astrophysics which couple a Monte Carlo representation of the radiation to a fluid model through interactions evaluated on a per-sample basis, yielding a Monte Carlo Radiation Hydrodynamics (MCRHD) scheme. This technique has received particular attention in the stellar physics community in \cite{haworthharries2012}, \cite{noebauer2012}, \cite{abdikamalov2012}, \cite{wollaeger2013}, and \cite{roth2014}, which have variously investigated extensions such as implicit methods and interfacing the Monte Carlo representation with a continuum approximation in regions of large optical depth and/or large ratio of radiation to gas pressure, where the unadorned Monte Carlo technique fails. MCRHD schemes have also been implemented for studying star formation in \cite{harries2015}, and have been used to model Compton cooling of accretion disks around black holes in flat space by \cite{ghosh2011} and \cite{garain2012}.  Monte Carlo techniques are particularly attractive for GRRMHD because they are algorithmically simple, naturally incorporate frequency dependence (useful for treating Compton scattering) and the potentially complicated angular dependence expected in an optically thin regime, and are easily modified to include special and general relativistic effects.  

In what follows we develop a scheme for GRRMHD called \bhl~that is designed to model accretion flows with modest to low optical depth. \bhl~couples two existing schemes: the GRMHD code \harm\footnote{Freely available; {\tt http://rainman.astro.illinois.edu/codelib/codes/harm/harm.tgz}}~\citep{gammie2003}, and the Monte Carlo radiative transport scheme \grm\footnote{Freely available; {\tt http://rainman.astro.illinois.edu/codelib/codes/grmonty/grmonty.tgz}}~ \citep{dolence2009}. The paper is organized as follows: \S 2 recounts the governing equations as they are solved in \bhl; \S 3 describes the numerical method; \S 4 demonstrates that \bhl~converges on a set of test problems; \S 5 describes example applications to a radiating Bondi flow and an M87-like disk model; \S 6 concludes.  

\section{Basic Equations}

We adopt a physical model in which emission, absorption, and scattering of photons couple an ideal, magnetized fluid to the radiation field.   We consider the fluid and radiation sector in turn.  The basic equations are identical to those integrated in the \harm~code \citep{gammie2003} and in the \grm~code \citep{dolence2009}, but recounted here to define variables and expose physical assumptions.

\subsection{Fluid}
\label{sec:fluidderivation}

We assume particle number conservation, which in a coordinate basis is
\begin{equation}
\partial_{t} \left( \sqrt{-g} \rho_0 u^{t}\right) = -\partial_{i} \left( \sqrt{-g} \rho_0 u^{i}\right),
\label{eq:masscons}
\end{equation}
where $\rho_0$ is the comoving frame rest mass density and $u^{\mu}$ is the fluid four-velocity.

Energy and momentum conservation for the coupled fluid {\it and} radiation system are given by 
\begin{equation}
\left(T^{\mu}_{~\nu} + R^{\mu}_{~\nu}\right)_{;\mu} = 0,
\label{eq:rsets}
\end{equation}
\noindent where $T^{\mu}_{~\nu}$ is the magnetohydrodynamic stress-energy tensor, and $R^{\mu}_{~\nu}$ is the radiation stress-energy tensor (not to be confused with the Ricci tensor). In a coordinate basis, Eq. \ref{eq:rsets} becomes
\begin{equation}
\partial_{t} \left( \sqrt{-g} T^{t}_{~\nu}\right) = -\partial_{i} \left( \sqrt{-g} T^{i}_{~\nu}\right) +\sqrt{-g}T^{\kappa}_{~\lambda}\Gamma^{\lambda}_{~\nu \kappa} + \sqrt{-g} G_{\nu},
\label{eq:enmomcons}
\end{equation}
where the radiation four-force density
\begin{equation}
 G_{\nu} \equiv -R^{\mu}_{~\nu;\mu}.
 \end{equation}
 In the ideal MHD limit $u_\mu F^{\mu\nu} = 0$ ($F^{\mu\nu} \equiv$ electromagnetic field tensor), and one can show that
\begin{equation}
T^{\mu}_{~\nu} = (\rho_0 + u + P + b^2)u^{\mu} u_{\nu} + (P + \frac{1}{2}b^2) g^{\mu}_{~\nu}-b^{\mu} b_{\nu},
\end{equation}
where $P \equiv$ fluid pressure, $u \equiv$ fluid internal energy density, and $b^2 = b^{\mu} b_{\mu}$, with $b^{\mu}$ the magnetic field four-vector,
\begin{equation}
b^{\mu} \equiv \frac{1}{2} \epsilon^{\mu \nu \kappa \lambda}u_{\nu} F_{\lambda \kappa},
\end{equation}
and $\epsilon^{\mu \nu \kappa \lambda} \equiv -[{\mu \nu \kappa \lambda}]/\sqrt{-g}$ is the Levi-Civita tensor.

Evidently $b^{\mu} u_{\mu} = 0$, so $b^\mu$ has only three degrees of
freedom, expressed as $B^i \equiv~^*F^{it}$, where
\begin{equation}
^*F^{\mu\nu} = \frac{1}{2} \epsilon_{\mu \nu \kappa \lambda} F^{\kappa\lambda} 
= b^\mu u^\nu - b^\nu u^\mu.
\end{equation}
Then
\begin{equation}
b^t = B^i u^{\mu} g_{i \mu},
\end{equation}
\begin{equation}
b^i = \frac{B^i + b^t u^i}{u^t}.
\end{equation}
The magnetic field evolution is determined by 
\begin{equation}
\partial_t \left( \sqrt{-g} B^i\right) = \partial_j \left[ \sqrt{-g} \left( b^j u^i - b^i u^j\right)\right]
\label{eq:induction}
\end{equation}
subject to the no-monopoles constraint
\begin{equation}
\partial_i \left( \sqrt{-g} B^i\right) = 0.
\label{eq:nomonopoles}
\end{equation}
The equation of state is
\begin{equation}
P = (\gamma - 1)u.
\label{eq:eos}
\end{equation}
To summarize: the governing equations for the fluid evolution are equations \ref{eq:masscons}, \ref{eq:enmomcons}, and \ref{eq:induction}, together with \ref{eq:nomonopoles} and \ref{eq:eos}.

\subsection{Radiation}
\label{sec:transport}

The radiation field consists of photons with wave four-vector $k^\mu$ and momentum $p^\mu = \hbar k^\mu$.  The photons follow geodesics, with 
\begin{equation}
\frac{{\mathrm d} x^{\mu}}{{\mathrm d} \lambda} = k^{\mu}.
\label{eq:dxdlambda}
\end{equation}
and
\begin{equation}
\frac{{\mathrm d} k^{\mu}}{{\mathrm d} \lambda} = -\Gamma^{\lambda}_{~\mu \nu} k^{\mu} k^{\nu},
\label{eq:geodesic}
\end{equation}
where $\Gamma^{\lambda}_{~\mu \nu} $ is the connection and $\lambda$ is an affine parameter along the geodesic.  We assume that plasma dispersion effects are negligible, so photons travel on null geodesics, $k_\mu k^\mu = 0$.   The frequency of a photon in a frame with four-velocity $u^\mu$ is $\omega = -k^\mu u_\mu$ ($\nu \equiv \omega/(2 \pi)$).

In nonrelativistic radiative transfer one describes the radiation field with the specific intensity $I_\nu$ (here and throughout we ignore polarization), which is frame-dependent.  However, $I_\nu/\nu^3 \propto f_R$ where $f_R$  is the radiation distribution function
\begin{equation}
f_R(x^\mu, p_i) = \frac{\d N}{\d^3x \d^3p},
\end{equation}
where $\d^3p = \d p_1 \d p_2 \d p_3$.  Because $\d N$, $\d^3x p^t \gdet$, and $\d^3p / (p^t \gdet)$ are invariant, $f_R$ is also invariant.

The evolution of $f_R$ is given by the Boltzmann equation,
\begin{equation}
\frac{\D f_R}{\d \lambda} = C[f_R]
\end{equation}
where $\lambda$ is an affine parameter along a photon trajectory (geodesic) and $C[f_R]$ accounts for interactions with matter: emission, absorption, and scattering of photons.  The Liouville operator $\D/\d\lambda$ is a derivative along the photon trajectory in phase space.  

One can rewrite the Boltzmann equation as the radiative transfer equation,
\begin{equation}
\frac{\D}{\d \lambda}\left( \frac{I_{\nu}}{\nu^3}\right) = \left( \frac{\eta_{\nu}}{\nu^2}\right) - \left(\nu \chi_{\nu} \right)\left( \frac{I_{\nu}}{\nu^3}\right).
\label{eq:transfer}
\end{equation}
Here the extinction coefficient  
\begin{equation}
\chi_{\nu} \equiv \alpha_{\nu} + \sigma_{\nu},
\end{equation}
and the emission coefficient
\begin{equation}
\eta_{\nu} \equiv j_{\nu} + \eta^s_{\nu}(I_{\nu}),
\end{equation}
where $j_{\nu}$ is the fluid emissivity, $\eta^s_{\nu}(I_{\nu})$ is the scattering contribution to emissivity, $\sigma_{\nu}$ is the scattering extinction coefficient, and $\alpha_{\nu}$ is the absorption coefficient.
Each of the quantities in parentheses in (\ref{eq:transfer}) is invariant.

We neglect stimulated Compton scattering.  The ratio of stimulated to spontaneous scattering is the photon occupation number in the scattered state.   Models of highly sub-Eddington accretion onto supermassive black holes commonly feature: (1) relativistic electrons with $\Theta_e \equiv k T_e/(m_e c^2) > 1$, corresponding to a mean amplification factor after Compton scattering of $\approx 16 \Theta_e^2$; (2) a low frequency (millimeter or far-IR) peak in the spectrum at $\nu_{pk}$ where the synchrotron absorption optical depth is $\order(1)$.   The energetically important single scattering events therefore produce scattered photons with $\nu_{sc} \sim \nu_{pk} 16 \Theta_e^2$.  For moderate accretion rates (i.e. scattering depth $\tau_s<1$ for the disk), the photon occupation number at $\nu_{sc}$ is small, and so stimulated Compton scattering will be negligible.   

A consequence of our neglect of stimulated Compton scattering is that in a purely scattering medium the radiation field will approach a Wien (Boltzmann) distribution rather than a Bose-Einstein distribution.  We verify this in Section \ref{sec:comptoncoolingtest}.

To summarize: the governing equations for the radiation are  (\ref{eq:transfer}), (\ref{eq:geodesic}), and (\ref{eq:dxdlambda}), together with appropriate expressions for the emission, scattering, and absorption coefficients.

\subsection{Radiation-Fluid Interactions}

In this section we adopt units such that $c = 1$ unless otherwise stated. It is apparent from Equations (\ref{eq:enmomcons}) and (\ref{eq:transfer}) that the fluid acts on the radiation through extinction and emission coefficients $\chi_{\nu}$ and $\eta_{\nu}$.  The radiation acts on the fluid through the four-force density $G_{\mu}$.
We want to make these representations consistent.  Begin with the manifestly covariant expression
\begin{equation}
R^{\mu\nu} = \int \frac{\d^3p}{\gdet p^t} \,\, p^\mu p^\nu f_R.
\label{eq:Rmunuf}
\end{equation}
Then 
\begin{equation}
G^\mu = -\nabla_\nu \int \frac{\d^3p}{\gdet p^t} p^\mu p^\nu f_R =
-h\int \frac{\d^3p}{\gdet p^t} p^\mu \frac{\D f_R}{\d\lambda}.
\label{eq:Gmu}
\end{equation}
The last equality follows from an expansion of $\D/\d \lambda$ and an integration by parts over momentum space (Lindquist 1966).  

Using $f_R = I_\nu/(h^4 \nu^3)$, and equations (\ref{eq:transfer}) and (\ref{eq:Gmu}), 
\begin{equation}
G^\mu = \frac{1}{h^3}\int \frac{\d^3p}{\gdet p^t} p^\mu \left[ (\nu {\chi}_{\nu}) ({I}_{\nu}/\nu^3) - ({\eta}_{\nu}/\nu^2) \right]
\end{equation}
where $\nu$, $\chi_\nu$, $I_\nu,$ and $\eta_\nu$ are all evaluated in a frame with four-velocity $u^\mu$. 

$G^\mu$ can be evaluated in an orthonormal tetrad frame comoving with the fluid
\begin{equation}
e^\mu_{(a)}, \,\,\,\, e^\mu_{(t)} = u^\mu
\end{equation}
We will call this the ``fluid frame.''  In the fluid frame,
\begin{equation}
G_{{(a)}} =  \int {\mathrm d}\nu{\mathrm d}\Omega \left( {\chi}_{\nu} {I}_{\nu} - {\eta}_{\nu} \right) n_{(a)},
\end{equation}
where $n_{{(a)}} \equiv p_{{(a)}}/(h \nu)$.  Then
\begin{equation}
G^\mu = e^\mu_{(a)} G^{(a)}.
\end{equation}
which is manifestly consistent with energy-momentum gains and losses by the radiation field.

\section{Numerical Method}

\bhl~combines a second order flux-conservative ideal GRMHD integrator \citep{gammie2003} with a Monte Carlo scheme for radiation transport \citep{dolence2009} through radiation-fluid interactions into a fully explicit GRRMHD scheme that is second order in space and first order in time for smooth flows. In this work we restrict ourselves to one- and two-dimensional flows, although the scheme can be trivially generalized to three spatial dimensions.   

\subsection{Fluid Integration}
\label{sec:fluidintegration}

The fluid integrator in \bhl~is taken from \harm, a conservative second order shock-capturing scheme on a two-dimensional mesh with an arbitrary spacetime metric.  Here we give a brief summary of the method. Also, we adopt units such that $c = 1$, and for black holes we set $GM = 1$. 

The fluid sector in \bhl~updates a set of conserved variables ${\bf U}$:
\begin{equation}
{\bf U} = \sqrt{-g} \left( \rho_0 u^t, T^t_{~t}, T^t_{~i}, B^i\right),
\end{equation}
corresponding to the variables whose coordinate time derivatives are given in \S \ref{sec:fluidderivation}. These conserved variables are updated by fluxes ${\bf F}$:
\begin{equation}
{\bf F} = \sqrt{-g} \left(\rho_0 u^i, T^i_{~t}, T^i_{~j}, B^i \tilde{v}^j - B^j \tilde{v} ^i \right),
\end{equation}
which in turn are calculated from the primitive variables ${\bf P}$:
\begin{equation}
{\bf P} = \left( \rho_0, u, \tilde{v}^i, B^i\right),
\end{equation}
where
\begin{equation}
\tilde{v}^i = v^i + \frac{\gamma \beta^i}{\alpha},
\end{equation}
where $v^i = u^i/u^0$ is the fluid spatial 3-velocity, $\gamma = \sqrt{1+g_{ij}u^i u^j}$, $\alpha = \sqrt{-1/g^{00}}$ is the lapse, and $\beta^i = g^{0i} \alpha$ is the shift. Unlike $v^i$, $\tilde{v}^i$ ranges over $-\infty < \tilde{v}^i < \infty$.  In the Newtonian formulation all transformations between the nonrelativistic analogs of ${\bf U}$, ${\bf F}$, and ${\bf P}$ are analytic, but in the covariant formulation there is no general analytic form for ${\bf P} ({\bf U})$.

The fluid update each timestep maps ${\bf P}^n$ to its updated value ${\bf P}^{n+1}$ by updating the conserved variables. Beginning with ${\bf P}^n$, the scheme calculates ${\bf U}^n = {\bf U} ({\bf P}^n)$ and ${\bf F}^n =  {\bf F} ({\bf P}^n)$ via closed-form expressions, for ${\bf F}^n$ after a reconstruction step that estimates ${\bf P}^n$ at zone boundaries from values at zone centers. The update ${\bf U}^n \rightarrow {\bf U}^{n+1}$ over a timestep $\Delta t$ is given by
\begin{equation}
{\bf U}^{n+1} = {\bf U}^n+ \Delta t \left(-\frac{{\bf F}^{n+1/2}_{i+1,j}-{\bf F}^{n+1/2}_{i,j}}{\Delta x^1}- \frac{{\bf F}^{n+1/2}_{i,j+1}-{\bf F}^{n+1/2}_{i,j}}{\Delta x^2}+ \dot{\bf U}^{n+1/2}\right) ,
\end{equation}
where $\dot{\bf U}$ represents the source terms such as those associated with the spacetime connection, values at $n+1/2$ are estimated from a similar first-order step to  ${\bf U}^{n+1/2}$, and $i,j$ here denote spatial indices in $x^1$ and $x^2$, respectively.  This forms a second order, explicit timestepping scheme to $t + \Delta t$, and then ${\bf P}^{n+1}$ is found by numerically solving ${\bf U} ({\bf P}^{n+1}) = {\bf U}^{n+1}$ (see \citealt{noble2006} and \citealt{mignone2007}).

The fluxes ${\bf F} ({\bf P})$ are evaluated at zone faces using Local Lax Friedrichs fluxes.  Primitive variables on either side of the zone face are determined through slope-limited linear reconstruction.  We typically use the monotonized central limiter for reconstruction, but it is trivial to use higher order methods as well.

Naively differencing the induction equation (\ref{eq:induction}) will not preserve a numerical representation of the no-monopoles constraint (\ref{eq:nomonopoles}); the monopole density will undergo a random walk from zero with a step size determined by truncation error.  Unless directly controlled, the monopole density can grow quickly and corrupt the solution.   A variety of techniques for avoiding or removing magnetic monopoles exist; \bhl~employs the flux-interpolated constrained transport (flux-CT) scheme introduced by \cite{toth2000}. Although this introduces some additional diffusivity into the scheme, it is simple and effective. Details of the implementation are given in \cite{gammie2003}.

\subsection{Radiation Transport}
\label{sec:radiation}

\bhl~uses nearly the same Monte Carlo implementation as \grm, with a few important differences.
The Monte Carlo samples are referred to here as superphotons.  Each superphoton has a weight $w$ (the number of photons carried by the superphoton), a momentum $p_\mu$ ($p_\mu = \hbar k_\mu$ and $p_\mu p^\mu = 0$), and a position $x^\mu$.

The Monte Carlo representation of the the photon distribution function is
\begin{equation}
f_{R,MC} = \sum_k \, w_k \, \delta^3(x^i - x^i_k) \delta^3(p_j - p_{j,k})
\label{eq:frmc}
\end{equation}
where $\delta^3(x^i - x^i_k) = \delta(x^1 - x^1_k)\delta(x^2 - x^2_k) \delta(x^3 - x^3_k)$, etc.
The sum is taken over all photon samples in the model, labeled by the index $k$, and $w_k$ are the weights.  Like $f_R$, $f_{R,MC}$ is invariant because $w_k$, $\delta^3(x^i-x^i_k)/(\gdet p^t)$, and $\delta^3(p_j-p_{j,k}) \gdet p^t$ (with $p_j$ {\it covariant}) are all invariant.

Using equation (\ref{eq:Rmunuf}), the stress-energy tensor is
\begin{equation}
R^{\mu\nu} = \sum_k \frac{p_k^\mu p_k^\nu}{\gdet p_k^t} w_k \delta^3(x^i - x^i_k)
\end{equation}
This can be averaged over a three-volume $\Delta^3 x = \Delta x^1 \Delta x^2 \Delta x^3$ to obtain an estimate for $\bar{R}^{\mu\nu}$:
\begin{equation}
\bar{R}^{\mu\nu} \approx \frac{1}{\Delta^3x} \int d^3x R^{\mu\nu} = \frac{1}{\gdet \Delta^3 x }\sum_k \frac{p_k^\mu p_k^\nu}{p_k^t} w_k
\end{equation}
where now the sum is taken only over photons within the three-volume (zone) in question.  

\subsubsection{Initializing the radiation field}

How should one initialize $f_{R,MC}$?   In \bhl's target applications this question usually does not arise because $f_R$ relaxes rapidly to a quasi-equilibrium, so one can set $f_R = 0$ in the initial conditions.  In test problems, however, an accurate initial $f_R$ may be required.  In this case one wants to sample a set of photons in a single zone centered at $x_c$, that is, we want to sample $\Delta^3 x f_R(x_c)$. 

One strategy is to sample $f_R$ directly in a coordinate frame, using the invariance of $f_R$.  For example, if $f_R$ is thermal in the fluid frame, then the distribution function in any coordinate frame is $c^2B_\nu/(h^4 \nu^3)$, where $B_{\nu}$ is the Planck function, and $\nu = -u^{\mu} k_{\mu}/(2 \pi)$. 

A second strategy, which we adopt, is to sample $f_R$ in the fluid frame (comoving indices are denoted by parentheses).  Then we must take care: $\Delta^3 x f_R(x_c)$ is not invariant, because the volume element $\Delta^3 x$ is not invariant.   But $\Delta^3 x \gdet p^t$ {\em is} invariant, so $(\Delta^3 x/\Delta^3 x') = \gdet p^t/p^{(t)}$, where $\Delta^3 x'$ is a fluid frame volume element, and $\gdet = 1$ in the fluid frame.   Then $\Delta^3 x f_R(x_c) = (\Delta^3 x/\Delta^3 x') \Delta^3 x' f_R = (\gdet p^t/p^{(t)} )\Delta^3 x' f_R$.  This suggests that we can sample $\Delta^3 x' f_R$ in the fluid frame and then multiply the photon number $\d N$ by the corresponding $\gdet p^t/p^{(t)}$ to obtain a fair sample of $\Delta^3 x f_R$ in the coordinate frame.

This second strategy can be described more explicitly in terms of the Monte Carlo samples as follows.  A list of photons in a single zone is obtained by taking
\begin{equation}
\int_{\Delta^3x}\, d^3x \, f_{R,MC} = \sum_k \, w_k \, \delta^3(p_j - p_{j,k}) 
\end{equation}
where the sum is over photon samples in a single zone.  This is not invariant because $\delta^3(p_j - p_{j,k})$ is not invariant, so one must take care in sampling $p_{j,k}$ and $w_k$.   Suppose we sample $f_R\Delta^3 x'$ in the tetrad frame.   This gives us a list of weights and momenta.   We can transform back to the fluid frame using the invariance of $\gdet p^t \delta(p_j-p_{j,k})$, so that each $w_k\delta(p_j-p_{j,k})$ in the tetrad frame becomes $w_k \gdet (p^{t}/p^{(t)}) \delta(p_j-p_{j,k})$ in the coordinate frame.  We can therefore obtain a fair sample by adjusting the weights by a factor of $\gdet p^{t}/p^{(t)}$ in transforming from the fluid frame to the coordinate frame.

\subsubsection{Geodesic integration}

The position and wavevector of each superphoton is evolved individually by integrating Equations (\ref{eq:dxdlambda}) and (\ref{eq:geodesic}) numerically.  Since evaluation of Christoffel symbols is costly, it is sensible to minimize the number of evaluations per timestep.   

We use the Verlet algorithm, a second order method that requires only one evaluation of the connection (number of evaluations is typically the order of the scheme). The algorithm as used in \bhl~is identical to that used in \grm~and is described explicitly in \cite{dolence2009}.

The Verlet method may be applied iteratively {\it without re-evaluating the Christoffel symbols}. For a fractional tolerance of $10^{-3}$ and the timesteps (corresponding to the $\Delta \lambda$) taken in \bhl, the scheme always converges. Although as of this writing we integrate all four components of $k^{\mu}$, one could potentially integrate three components and use $k^{\mu} k_{\mu} = 0$ to evaluate the fourth, suppressing numerical errors and computational expense by a factor of $4/3$.

\subsubsection{Units}

The radiation sector uses cgs units, except that photon wavevector components are measured in units of the electron rest mass energy.  We therefore have to convert between units in the fluid sector and units in the radiation sector.  For black hole spacetimes, this is done by choosing a cgs value for the fluid length unit and the fluid time unit, here
\begin{equation}
\lunit = \frac{G M}{c^2},
\end{equation}
and
\begin{equation}
\tunit= \frac{G M}{c^3},
\end{equation}
respectively. We also need a mass unit.  Notice that the mass unit is {\em not} provided by the black hole mass in the test fluid ($\rho \lunit^3 \ll M$) approximation used here.  Instead we must scale the density, or equivalently the mass accretion rate, by choosing a cgs value for the mass unit $\munit$.  Then, e.g., $\rho_{\mathrm{CGS}} =  \rho_{\mathrm{FLUID}} \munit/\lunit^3$.

The components of the code photon wavevector $k^{\mu}$ are measured in units of $m_e c^2$.  One might then be concerned about consistency between the transfer equation and the geodesic equation.  The only condition for consistency is that the differential optical depth $d\tau_\nu = (\nu \kappa_\nu ) d\lambda$, which in turn requires that the $\nu d\lambda = ds$, i.e. that the units used in defining $\nu$ and $d\lambda$ be consistent and that the correct conversion be made from fluid sector units to cgs.  In practice, then, we evaluate $\nu \kappa_\nu$ in cgs units in the fluid frame and set $d\tau_\nu = \mathcal{N} (\nu \kappa_\nu) d\lambda$, with $\mathcal{N} \equiv h\lunit/(m_e c^2)$.

\subsubsection{Superphoton Weighting}

The passive Monte Carlo code \grm~is designed to maximize the signal to noise in the final spectrum, which is measured in logarithmic intervals in frequency at spatial infinity.    The optimum allocation of weights would then place equal numbers of superphotons in each bin in $\log\nu$.  This requires an estimate of the final spectrum;  \grm~estimates the final spectrum by integrating over the simulation volume, assuming the flow is optically thin at all frequencies, neglecting gravitational redshift and Doppler shift, and setting the weights accordingly.

In \bhl, by contrast, the weights should be designed to minimize errors in the dynamical evolution, i.e.~in $G_{\mu}$.  The momentum and energy exchange associated with each radiation-fluid interaction is proportional to $w h\nu$, where $\nu$ is the fluid frame frequency.  This suggests that we should distribute energy uniformly among superphotons (as in \citealt{abbott1985}) to minimize the interaction noise, and thus set $w \propto 1/\nu$.  This is not generally possible because the four-velocity fluctuates across the simulation domain, but we will not do too badly if we ignore Doppler shift and gravitational redshift and thus set $w \propto 1/\nu$ when sampling the emissivity.

\bhl's constant-energy-per-superphoton weighting scheme limits spectral resolution at low and high frequencies where the specific energy density is small prior to scattering.  These parts of the spectrum have little impact on the dynamical evolution, however, and higher quality spectra can be extracted in post-processing using \grm.

\subsubsection{Emissivity}
\label{sec:emissivity}

At each timestep we sample the emissivity in the invariant four-volume $\gdet \Delta t \Delta^3 x$ of each zone based on the fluid values at the half-step.   It is easiest to sample the fluid emission in a comoving tetrad, where we have a simple expression for the emissivity.   

The emissivity is
\begin{equation}
j_\nu = \frac{1}{\gdet}\frac{dE}{d^3x dt d\nu d\Omega}.
\end{equation}
Using $dE = h\nu dN$ where $N$ is the number of photons, we can write
\begin{equation}
\int d\Omega\, j_\nu = \frac{h}{\gdet} \frac{dN}{d^3x dt d\log\nu}.
\end{equation}
Since $dN = w dN_s$ where $N_s$ is the number of superphotons, we can then write for the number of superphotons produced per logarithmic interval in a single zone with volume $\Delta^3x$:
\begin{equation}
 \frac{dN_s}{d\log \nu} = \sqrt{-g}\Delta t \Delta^3 x \frac{1}{h w(\nu)} \int {\mathrm d}\Omega \, j_{\nu}.
\label{eq:freqdist}
\end{equation}
In writing (\ref{eq:freqdist}) we have made use of the invariance of $\gdet d^3x dt$.  Here $w(\nu) \propto 1/\nu$, and the constant of proportionality is set dynamically to keep the number of superphotons in the computational domain approximately constant.

\bhl~samples equation \ref{eq:freqdist} between minimum and maximum frequencies $\nu_{\textrm{min}}$ and $\nu_{\textrm{max}}$, where the limits are set so that $\nu j_{\nu}$ is small outside this region in frequency space.  It then uses a rejection scheme, sampling a uniform distribution in $\log\nu_{min} < \log\nu < \log\nu_{max}$ and a uniform distribution in $0 < r < 
({\mathrm d}N_s/{\mathrm d} \log \nu)_{\textrm{max}}$, where $({\mathrm d}N_s/{\mathrm d} \log \nu)_{\textrm{max}}$ is the maximum of Equation \ref{eq:freqdist}.  A sample is rejected when $r > {\mathrm d}N_s/{\mathrm d} \log \nu$.

The angular distribution of photons is also sampled by rejection.  The photon direction is given by $(\theta,\phi)$, where $\theta$ is the angle between the magnetic field and photon direction in the fluid frame and $\phi$ is the corresponding azimuthal angle.  \bhl~samples a uniform distribution in $0 \le \cos\theta < 1$, and a uniform distribution in $0 \le r < 1$.  A sample is rejected if $r > j_{\nu}(\theta)/j_{\nu,{\textrm{max}}}$.  It then samples a uniform distribution in $0 \le \phi < 2\pi$.  To ensure that the net force due to emission in the fluid frame is zero to machine precision, photons are generated in pairs. Thus, a second photon is generated with the same frequency and $\cos\theta' = -\cos\theta$ and $\phi' = \phi + \pi$. In the fluid frame, $k^{(t)} = \omega$, $k^{(x)} = \omega\sin\theta\cos\phi$, $k^{(y)} = \omega\sin\theta\sin\phi$, and $k^{(z)} = \omega\cos\theta$, where $e^{(z)}$ is parallel to the magnetic field.  Once we have a superphoton sample in the comoving frame it is transformed to the coordinate frame using a pre-constructed orthonormal tetrad $e^\mu_{(a)}$.  The superphoton $x^\mu$ is set to the zone center to avoid additional orthonormal tetrad construction.\footnote{Because photons are created at zone centers, our scheme will fail when individual zones become optically thick.  Should this become a problem the scheme can be modified so that new superphotons are distributed within a zone.}

Sampling is a subdominant computational expense in \bhl, so although one could develop more efficient sampling schemes, a simple rejection scheme is adequate.

Four-momentum is locally conserved and so superphoton emission implies a back-reaction on the emitting fluid. A pair of emitted superphotons with wavevectors ${k}^{\mu}_1$, ${k}^{\mu}_2$ correspond to a change in four-momentum $\Delta p^{\mu}$ (in fluid code units):
\begin{equation}
\Delta p^{\mu} = \mathcal{P} w \left( k_1^{\mu} + k_2^{\mu}\right),
\end{equation}
where $\mathcal{P} = m_e/\mathcal{M}$, which in turn specifies the contribution to the four-force density $\Delta{G}_{\mu}$:
\begin{equation}
\Delta G_{\mu} = -\frac{1}{\sqrt{-g}\Delta^3 x\Delta t} \Delta p_{\mu},
\end{equation}
where all geometric quantities are evaluated at zone centers.  Because photons are emitted in pairs, the spatial components of $k^\mu$ cancel in the fluid frame, and $\delta p_\mu \propto u_\mu$.  

\subsubsection{Absorption}

\grm~treats absorption deterministically by continuously decrementing $w$ along a ray.  A similar deterministic procedure has been shown to suppress noise in Monte Carlo radiation hydro schemes (e.g.~\citealt{noebauer2012}) in flat space.  However, formulating such a scheme in general relativity, where photons move along geodesics, is more complicated because the photons follow curved trajectories through each zone.

In \bhl~we treat absorption probabilistically.  While stepping a superphoton by $\Delta\lambda$ along a geodesic, the incremental optical depth to absorption $\Delta\tau_a = \mathcal{N} \kappa_{\nu,abs} \nu \Delta \lambda$.  Here $\nu \kappa_{\nu,abs}$ is the invariant absorption coefficient, evaluated in the fluid frame and interpolated to $x^\mu$.  An absorption occurs if 
\begin{equation}
\Delta \tau_a > - \log r_a,
\label{eqn:abscondition}
\end{equation}
where $0 < r_a < 1$ is sampled uniformly; the absorption occurs at $\Delta\tau_a = -\log r_a$.  To process the event we push the superphoton back, $\lambda \rightarrow \lambda+\Delta \lambda(\log r_a /\Delta \tau_a)$, put the superphoton four-momentum into the fluid at that location, and annihilate the superphoton.

The four-momentum change in the fluid $\Delta {p}^{\mu}$ due to absorption of a superphoton with wavevector ${k}^{\mu}$ is 
\begin{equation}
\Delta p ^{\mu} = \mathcal{P} w k^{\mu}.
\end{equation} 
This can be expressed as a contribution to the radiation four-force density $\Delta G_{\mu}$:
\begin{equation}
\Delta G_{\mu} = \frac{\Delta p_{\mu}}{\gdet \Delta^3x \Delta t},
\end{equation}
where $\gdet$ is evaluated at the zone center.

\subsubsection{Scattering}
\label{sec:scattering}

We treat scattering probabilistically in \bhl, as in \grm.  Scattering is similar to absorption, i.e.~scattering occurs when 
\begin{equation}
\Delta \tau_s > - (\log r_s)/b_s,
\label{eqn:scattcondition}
\end{equation}
where $0 < r_s < 1$ is sampled uniformly, $\Delta\tau_s = \mathcal{N} \kappa_{\nu,s} \nu \Delta \lambda$, and $\nu\kappa_{\nu,s}$ is the invariant scattering opacity.   Because scattering events are rare but energetically important we have introduced a bias parameter $b_s > 1$ to enhance the probability of sampling scattering events.  To process the event, we push the photon back along the geodesic from $\lambda + \Delta \lambda$ to $\lambda+\Delta \lambda \log r_s / (b_s \Delta \tau_s)$.  To preserve photon number a scattered superphoton is created with weight $w_s = w/b_s$ and the original superphoton has weight set to $w' = w - w_s$.  

In general, a superphoton is subject to both absorption and scattering simultaneously. In a deterministic treatment, the code must dynamically choose which process, if any, to apply to the superphoton. To handle this in an unbiased manner, for each photon we, assuming that at least one of the inequalities Eqns. \ref{eqn:abscondition} and \ref{eqn:scattcondition} has been satisfied, choose which interaction to process according to a similar weighted sampling. That is, for 
\begin{equation}
\frac{-\log r_a}{\Delta \tau_a} < \frac{-\log r_s}{b_s \Delta \tau_s},
\end{equation}
the absorption interaction is chosen; else, scattering is chosen. With this method, large optical depth or bias in one interaction will not serve to decrease the physical effect of the other interaction, although it will increase the number of superphotons required to resolve both interactions simultaneously.

How should we set $b_s$?  Most of our models have $\tau_s \ll 1$, so only $\approx \tau_s$ superphotons would produce scattering events if $b_s = 1$, and the energy per superphoton would increase by the mean amplification factor $A \approx 1 + 4\Theta_e + 16 \Theta_e^2$.\footnote{This approximate expression overestimates $A$ by $16\%$ at $\Theta_e \approx 1/2$.  A better estimate, which underestimates $A$ by $4\%$ at $\Theta_e \approx 0.02$, is $A-1 \approx 4\Theta_e - 2\Theta_e^{3/2} + 16\Theta_e^2$.}     This suggests that we should set $b_s \sim A$ to maintain constant energy per superphoton.  There are two failure modes to be avoided, however.  First, if $b_s \gtrsim 1/\tau_s$ then each superphoton will scatter more than once and the number of superphotons on the grid will grow exponentially.   Second, if $A w h\nu$ is larger than the total internal energy in the zone where the scattering occurs then the zone energy will be negative after scattering (more on this in the next subsection).  Together, this suggests that we set
\begin{equation}
 b_s = A \operatorname{MAX} \left( {\mathcal{C}}, \frac{ w h \nu}{u \sqrt{-g} \Delta^3 x}\right).
\end{equation}
Here ${\mathcal C}$ depends only on $t$ and is dynamically adjusted to control the number of scattered superphotons in the simulation. The requirement $b_s \tau_s \sim 1$ is equivalent to each superphoton scattering once between emission and escape through the boundary (neglecting absorption). Over a timestep $\Delta t$, one can estimate the number of photons which escape through the boundaries of a domain with linear dimension $L$ as $N c \Delta t / L$, where $N$ is the desired total number of superphotons (which sets the weights for emission as described previously). To enforce the requirement that each superphoton scatter once, ${\mathcal C}$ is calculated dynamically as the ratio of this estimate to the real number of scattering events per timestep, averaged over some timescale.

Each scattered superphoton is generated from an incident superphoton wavevector $k^{\mu}$ as follows.  The four-momentum $p^{\mu}$ of the scattering electron is sampled from a thermal (Maxwell-J\"uttner) distribution according to the procedure described in \cite{canfield1987}.  The scattered superphoton wavevector $k^{\mu}_s$ is sampled from the Klein-Nishina differential scattering cross section in the rest frame of the scattering electron and boosted to the fluid frame and then transformed to the coordinate frame.  It is then assigned a weight and entered in the list of active superphotons.

Each scattering event generates a change in fluid four-momentum,
\begin{equation}
\Delta p^{\mu} = \mathcal{P} \frac{w}{b_s}\left(k^{\mu} - k_s^{\mu} \right)
\end{equation}
and a corresponding contribution to the fluid through the four-force density $G_{\mu}$,
\begin{equation}
\Delta G_{\mu} = \frac{\Delta p_{\mu}}{\gdet \Delta^3x \Delta t}
\end{equation}
where $\gdet$ is evaluated at the zone center.

\subsection{Radiation Force in Fluid Evolution}

The radiation force is treated in an operator-split fashion.  The fluid integrator initially updates the conserved variables ${\bf U}$ from step $n$ to $n+1$ over the entire timestep $\Delta t$ without radiation, i.e.~it performs ${\bf U}^n\rightarrow{\bf U}^{n+1'}$ as described in Section \ref{sec:fluidintegration}. This fluid integration generates half-step fluid primitive variables ${\bf P}^{n + 1/2}$; these values are sent to the radiation sector and used to evaluate the total radiation four-force density $G_{\mu}({\bf P}^{n + 1/2})$ for each zone. The fluid integrator then updates the fluid variables with the radiation interaction ${\bf U}^{n+1'} \rightarrow {\bf U}^{n+1}$ by considering only the radiation contribution to the conserved energy and momentum variables:
\begin{equation}
\left( \sqrt{-g} T^{t}_{~\nu}\right)^{n+1} = \left( \sqrt{-g} T^{t}_{~\nu}\right)^{n+1'} + \Delta t \sqrt{-g}G_{\nu}.
\end{equation}
${\bf U}^{n+1}$ then are the final conserved fluid variables at the $(n+1)^{\mathrm{th}}$ step.   The evolution is therefore first order in time.

The evolution is explicit and the radiation and fluid share a common timestep $\Delta t$, which we set to the minimum grid zone light crossing time $\approx \Delta x/c$, where $\Delta x$ is a characteristic zone lengthscale.  As is well known, the radiation source terms are stiff when the timescale for exchange of energy-momentum between the fluid and radiation is smaller than a timestep.  The cooling time $\tau_{cool} \equiv  u/\Lambda$, where $\Lambda \equiv$ cooling rate per unit volume, $= u_\mu G^\mu \sim u_{rad} c/\lambda_{mfp}$, where $\lambda_{mfp}$ is a suitably frequency-averaged absorption mean free path and $u_{rad} = R^{\mu\nu} u_\mu u_\nu$ is the radiation energy density in the fluid frame (one can perform a similar estimate for Compton cooling).  Thus the source term is stiff if $u/\Lambda < \Delta x/c$ or $(u_{rad}/u)(\Delta x/\lambda_{mfp}) > 1$, or when the optical depth across a zone exceeds $u/u_{rad}$.

For our scheme we must also consider robustness in the presence of Monte Carlo noise.  Even if $\tau_{cool}/\Delta t > 1$ the cooling rate may fluctuate upward so that a zone loses all its thermal energy in a single timestep.   This can happen if $u/\<\Lambda^2\>^{1/2} \lesssim \Delta x/c$.  This will differ from the usual stiffness condition only when the number of absorption events per timestep in a zone is small compared to one.  The condition for robustness against this failure mode, ``supercooling,'' where a single photon causes the zone to lose all its internal energy, is that $w h\nu < u \Delta^3 x$ (where we have left out geometric factors).

Where, then, will \bhl~fail?  The radiation force source terms are stiff when the optical depth across a single zone exceeds $u/u_{rad}$.  For black hole accretion applications we expect this only for models in the high accretion rate regime, $\dot{M} \gtrsim 10^{-3}\dot{M}_{Edd}$, although the precise condition will depend on details of the evolution and the numerical setup.  This problem could be remedied by using an implicit update, but Monte Carlo is probably not the optimal method for studying this regime anyway.  The supercooling problem is more relevant for our target application to intermediate accretion rate black holes, and arises if the internal energy content of a zone is small compared to the typical superphoton energy.  This can occur in low density regions over the poles of the black hole, but the fluid evolution is inaccurate there in any case (because the truncation error in internal energy is dominated by the magnetic field evolution, to which it is coupled via the total energy density), and negative internal energies are dealt with by {\tt harm}'s floor routines, resulting in a small nonconservation of energy.

\subsection{Parallelization}

\bhl~is a hybrid MPI/OpenMP scheme in which a single node handles the fluid integration, multiple nodes evolve the radiation, and a single additional node acts as gatekeeper between the fluid and radiation sectors.  
During each timestep, the only exchanges are an array of radiation four-force densities to the fluid sector and an array of fluid variables to the radiation sector via the gatekeeper node. The gatekeeper node distributes the fluid variables to all radiation nodes, and reduces the four-force density contributions from each radiation node. 

We evolve radiation on each node independently from other radiation nodes. After globally scaling the emission weights and scattering bias to yield approximately the desired number of superphotons at saturation, the code samples emission events on each radiation node, each of which has access to the entire array of fluid variables and maintains its own set of superphoton samples. Emission, absorption, and scattering events generate a four-force density contribution. At the end of every timestep, these contributions are reduced by the primary radiation node over MPI. Each radiation node is individually parallelized under OpenMP, further dividing the superphoton calculations across individual compute cores.  We parallelize the main compute loops for the fluid sector with OpenMP, which enables completion in a reasonable clock time for an axisymmetric calculation.

\subsection{Implementation Details}
\label{sec:miscellaneous}

In attempting to describe our numerical implementation in a coherent narrative we have omitted certain secondary topics, which we now collect here.
\begin{itemize}
\item The radiation sector in \bhl~makes extensive use of random numbers. We use the Mersenne Twister algorithm from the GNU Scientific Library, with a different random seed for each MPI node and each OpenMP thread.

\item In axisymmetric disk calculations, we implement a form of static mesh refinement by using modified Kerr-Schild (MKS) coordinates $\{ t, x^1, x^2\}$. $x^1$ and $x^2$ are related to the Kerr-Schild $r, \theta$ by $r = \exp \left( x^1\right) + r_0$ and $\theta = \pi x^2 + ((1-h_s)/2) \sin(2 \pi x^2 )$, where $r_0 \in [0, \infty)$ and $h_s \in (0, 1]$ are free parameters.

\item Truncation error in the geodesic integration causes $k^\mu$ to drift off the lightcone (this is a consequence of our decision to integrate all four components).  We destroy superphotons with negative frequency in the fluid frame; for torus runs as in Section \ref{sec:torus}, we find $\sim 1.1 \times10^{-6}$ destructions per geodesic update.  This problem does not occur in Cartesian coordinates in Minkowski space.
\item Scattered superphotons may scatter any number of additional times during the same timestep, provided sufficient optical depth to do so.
\item \bhl~does not conserve momentum and energy to machine precision because of truncation error in the geodesic integrator. However, for the integrator tolerance and typical superphoton resolutions this is not significant (for torus runs as in Section \ref{sec:torus}). At some time, the average fractional error in energy relative to the initial energy at emission is $\sim \mathrm{few}\times10^{-7}$; if this were to become a leading source of error, increasing the integrator tolerance is not a significant expense.
\end{itemize}

\section{Test Suite}

We have developed a suite of test problems for \bhl.  Since the fluid and radiation sectors of \bhl~ use well tested codes, we focus on problems  with coupling between the two sectors.  Good test problems are hard to find, since there are few known exact  solutions to the equations of radiation MHD in either Newtonian or relativistic contexts.  We substitute approximate solutions to the full equations, such as some of the shocks we consider below.  We do not consider pure transport tests that are trivially satisfied by a Monte Carlo scheme, such as shadow tests, expanding pulses, and dynamic diffusion\footnote{Such tests can be performed with the freely available \grm~code.}.

\subsection{Optically Thin Cooling}
\label{sec:cooling}

We consider the temperature evolution of an optically thin, radiating, stationary, ideal, and homogeneous gas initially at temperature $T_0$. The gas obeys a $\gamma$-law equation of state i.e.~$p = (\gamma - 1) u$.  The density and velocity of the gas are fixed; only the temperature is allowed to evolve.   The bremsstrahlung-like emissivity is
\begin{equation}
j_{\nu} = N n^2 T^{-1/2}\exp{\left(-h \nu / k_B T\right)},
\label{eqn:brememiss}
\end{equation}
\noindent where $N = 5.4\times10^{-39}~\mathrm{cm^3~K^{1/2}~s^{-1}~Sr^{-1}~Hz^{-1}}$  is a constant. The associated cooling rate $\Lambda$ is given by
\begin{equation}
\Lambda = -\frac{\mathrm{d}u}{\mathrm{d}t} = \int\mathrm{d}\nu\mathrm{d}\Omega j_{\nu} = 4 \pi \frac{k_B n^2 N}{h}T^{1/2},
\end{equation}
which implies a temperature evolution
\begin{equation}
T(t) = T_0\left(1-\frac{t}{t_f} \right) ^2
\end{equation}
\noindent valid from $t_i = 0$ to $t_f =h T_0^{1/2}/((\gamma - 1)\pi N n)$, the time at which the temperature of the fluid reaches zero. 

\begin{figure}[h]
  \caption{Optically thin cooling of one static fluid zone. Approximately $5\times10^8$ superphotons were created.}
  \centering
   \plotone{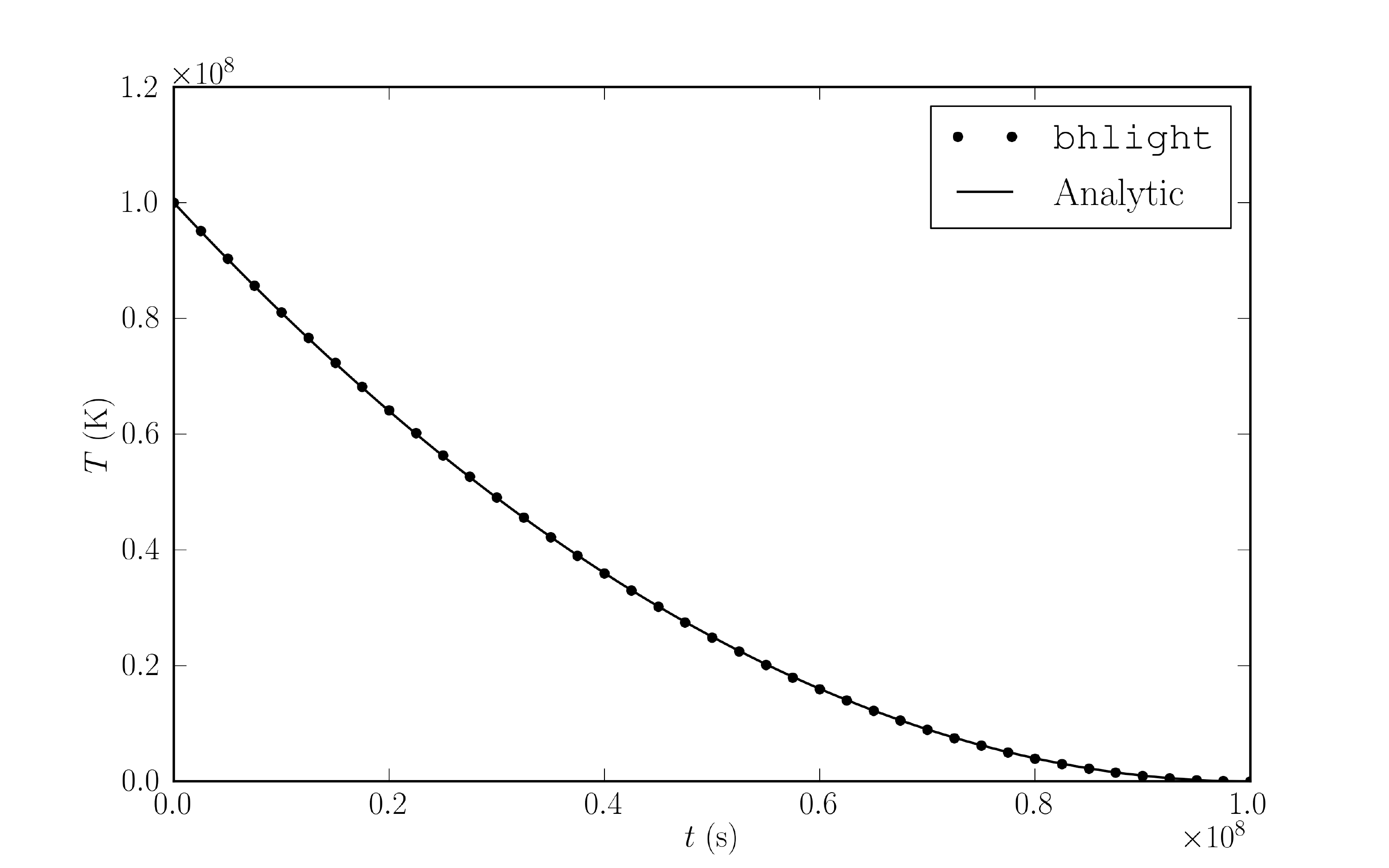}
\label{fig:opticallythinoverlay}
\end{figure}

\begin{figure}
  \caption{Convergence of the optically thin cooling test. $N_s$ is directly proportional to the number of superphotons created.}
  \centering
  \plotone{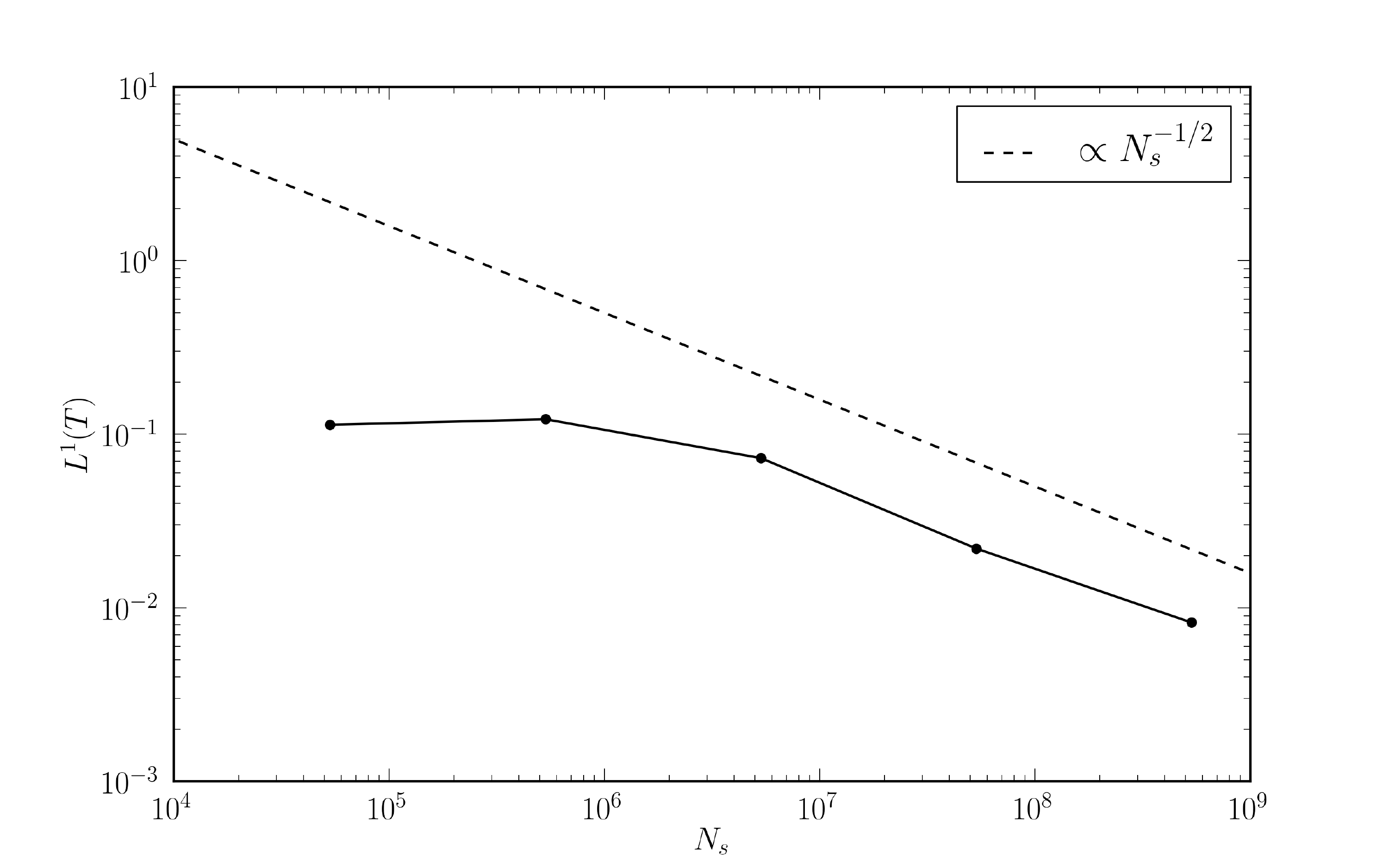}
\label{fig:opticallythinconvergence}
\end{figure}

For this realization we choose $\gamma = 5/3$, $t_f = 10^8~\mathrm{s}$, and $T_0 = 10^8~\mathrm{K}$. Figure \ref{fig:opticallythinoverlay} shows the resulting numerical evolution plotted against the analytic solution.  Convergence in the $L^1$ norm,
\begin{equation}
L^1(f) \equiv \sum_i  \left| f_{\mathrm{num},i}-f_{\mathrm{reference},i}\right|
\end{equation}
is expected to scale as $N_s^{-1/2}$; this can be seen in Figure \ref{fig:opticallythinconvergence}, which is evaluated near $t = t_f$.  

 \subsection{Compton Cooling}
 \label{sec:comptoncoolingtest}

Consider a closed, one-zone model in which Compton scattering is the only permitted interaction between an ideal, $\gamma = 5/3$ gas initially at temperature $T_{g,i}$ and a swarm of photons all with initial frequency $\nu=\nu_0$. Fluid motion is suppressed; only the internal energy is allowed to evolve. The number of photons is conserved, and in thermal equilibrium $T_{g,f} = T_{r,f} = T_f$, the radiation approaches the Wien distribution, 
\begin{equation}
f_R = \frac{n_{\gamma}}{8 \pi} \left( \frac{c}{k T_f} \right)^3 \exp{(-h \nu /( k_B T_f))},
\end{equation}
where $n_{\gamma}$ is the number density of photons. This tests the scattering kernel and Compton heating and cooling of the fluid.

\begin{figure}
  \caption{Evolution of the Comptonization problem. The top panel shows radiation and gas temperatures approaching the analytic final temperature. The middle panel shows the initial and final $u_{\nu}$ for the radiation, along with the analytic result for the final state. The bottom panel shows the residuals for the final spectrum; the numerical spectrum is apparently unbiased even at frequencies with low sampling resolution (shaded regions).}
  \centering
    \plotone{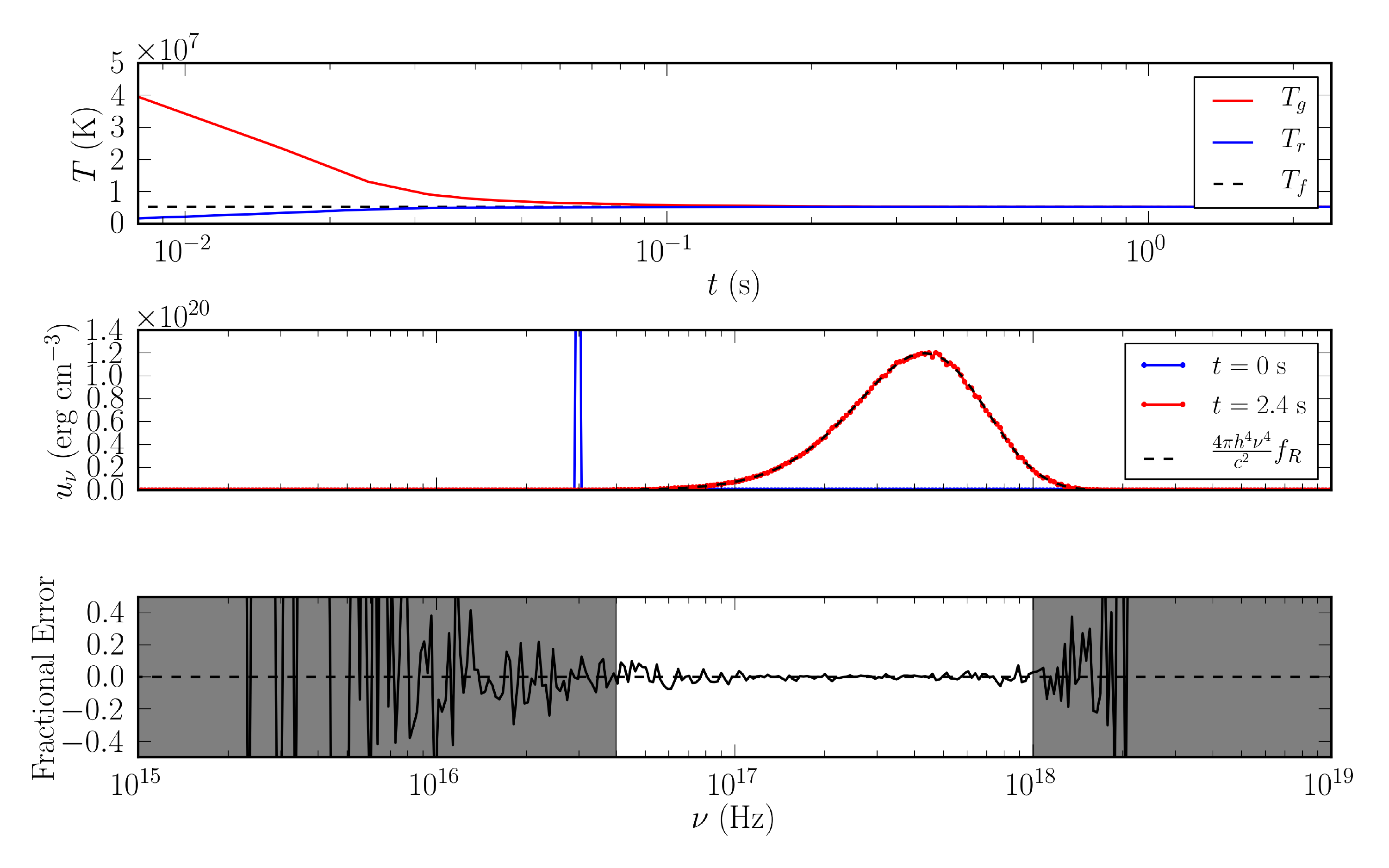}
\label{fig:comptonisationtriple}
\end{figure}

We set the electron number density $n = 2.5\times10^{17}$ cm$^{-3}$, $T_{g,i} = 5\times10^7~K$, $\nu_0 =  3\times10^{16}\mathrm{~Hz}$, and $n_{\gamma} = 2.38\times10^{18}\mathrm{~cm^{-3}}$. The characteristic (Compton) relaxation time is the photon mean free time $(n_e \sigma_T c)^{-1}$ divided by the fractional energy change per scattering, $\sim k T/(m_e c^2)$, yielding a Comptonization timescale $\simeq 0.02$ s. We can predict the final state using: (1) thermal equilibrium; (2) conservation of photon number; (3) that the final photon distribution is Boltzmann; (4) conservation of total energy.  This yields $T_f = 5.19 \times 10^6$ K. Figure \ref{fig:comptonisationtriple} shows fluid and gas temperatures equilibrating at the correct temperature on approximately the estimated timescale, and $u_\nu \equiv \d E / (\d^3 x \, \d \log \nu)$ plotted against the anticipated distribution along with associated residuals. Note that we define $T_r \equiv E_r /(3 n_{\gamma} k_B)$ ($E_r$ is the radiation energy density), which assumes a Wien distribution, and so this value is strictly valid only at late time.

\subsection{Linearized Transfer and Energy Equations}

\begin{figure}
  \caption{Dispersion relation for eigenmode of the transfer and energy equations as a function of optical depth per wavelength. Solid line shows analytic expectation, while points show \bhl\ results. At this resolution, the fractional error is $\approx 10^{-3}$.}
  \centering
    \plotone{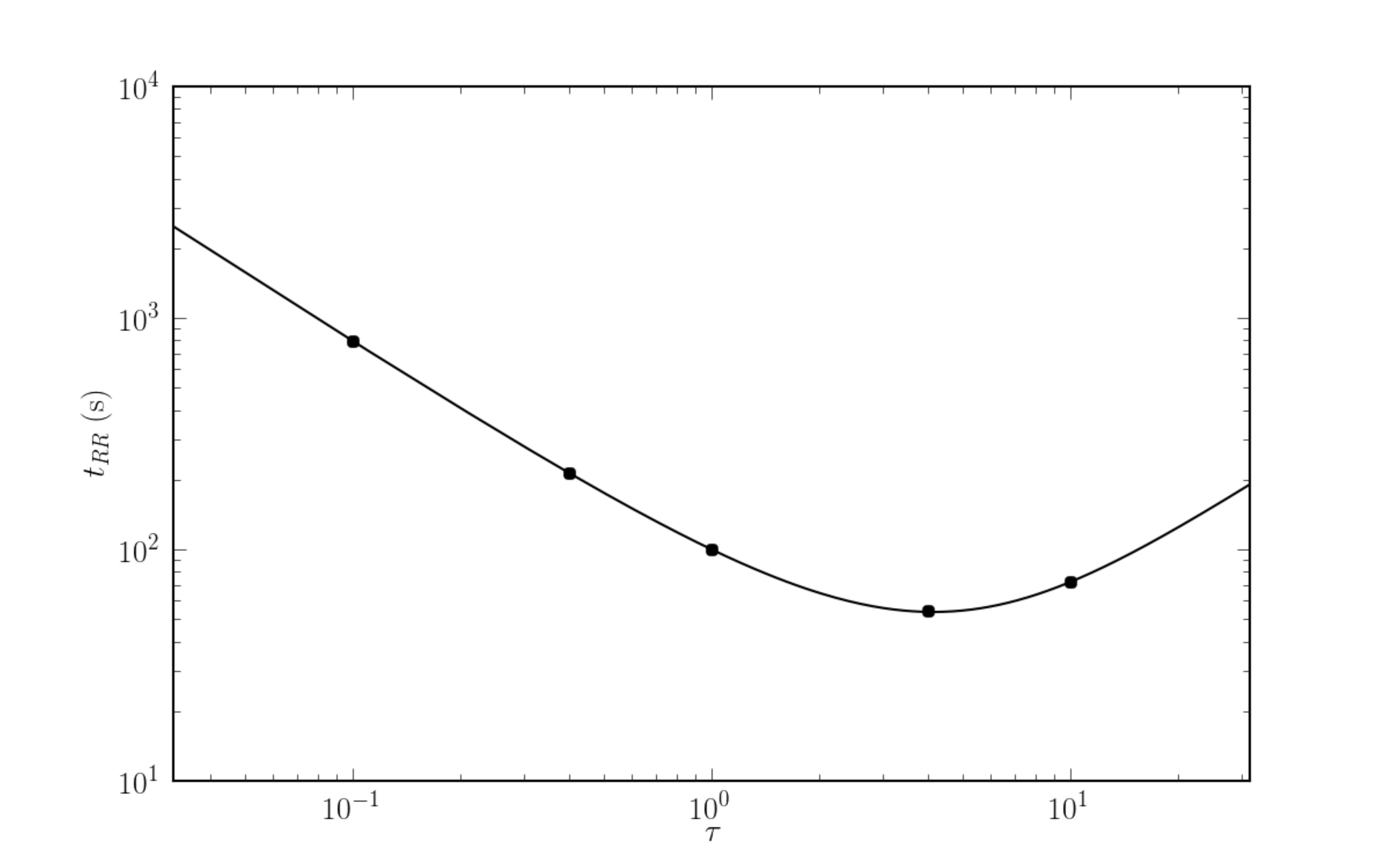}
\label{fig:dispersionrelation}
\end{figure}

As a test of the full transfer equation in 1D with gray absorption, consider a sinusoidal temperature perturbation in a static gas.  \cite{mihalasmihalas} show that in this case the full transfer equation plus gas energy equation admit damped solutions.  The eigenmode has perturbations $\propto \exp(i(k x - t/t_{RR}))$ for wavenumber $k$ and decay time $t_{RR}$.  The dispersion relation is
\begin{equation}
t_{RR}(k) = \left[ \frac{16 \sigma \kappa T_0^3}{\rho c_v} \left(1-\frac{\alpha_0}{k}\cot^{-1}\left( \frac{\alpha_0}{k}\right) \right) \right]^{-1},
\label{eqn:dispersionrelation}
\end{equation}
where $T_0$ is the mean temperature, $\rho$ is the material density, $\alpha_0$ is the frequency-integrated extinction coefficient, and $c_v$ is the specific heat capacity. We simulate this problem in \bhl~by initializing one wavelength in a 1D box in local radiative equilibrium with 64 grid zones and periodic boundary conditions. The amplitude of the initial perturbation is $0.05 T_0$. We evolve this system for a variety of optical depths per wavelength $\tau$ by varying $\alpha_0$. We obtain a decay time from the amplitude of the best fit sinusoid at $t = t_{RR}$. We find good agreement with Equation \ref{eqn:dispersionrelation} in both optically thin and optically thick regimes, as shown in Figure \ref{fig:dispersionrelation}.

\begin{figure}
  \caption{Convergence for the linear mode of the transfer and energy equations, with $N_s$ directly proportional to the number of superphotons.}
  \centering
    \plotone{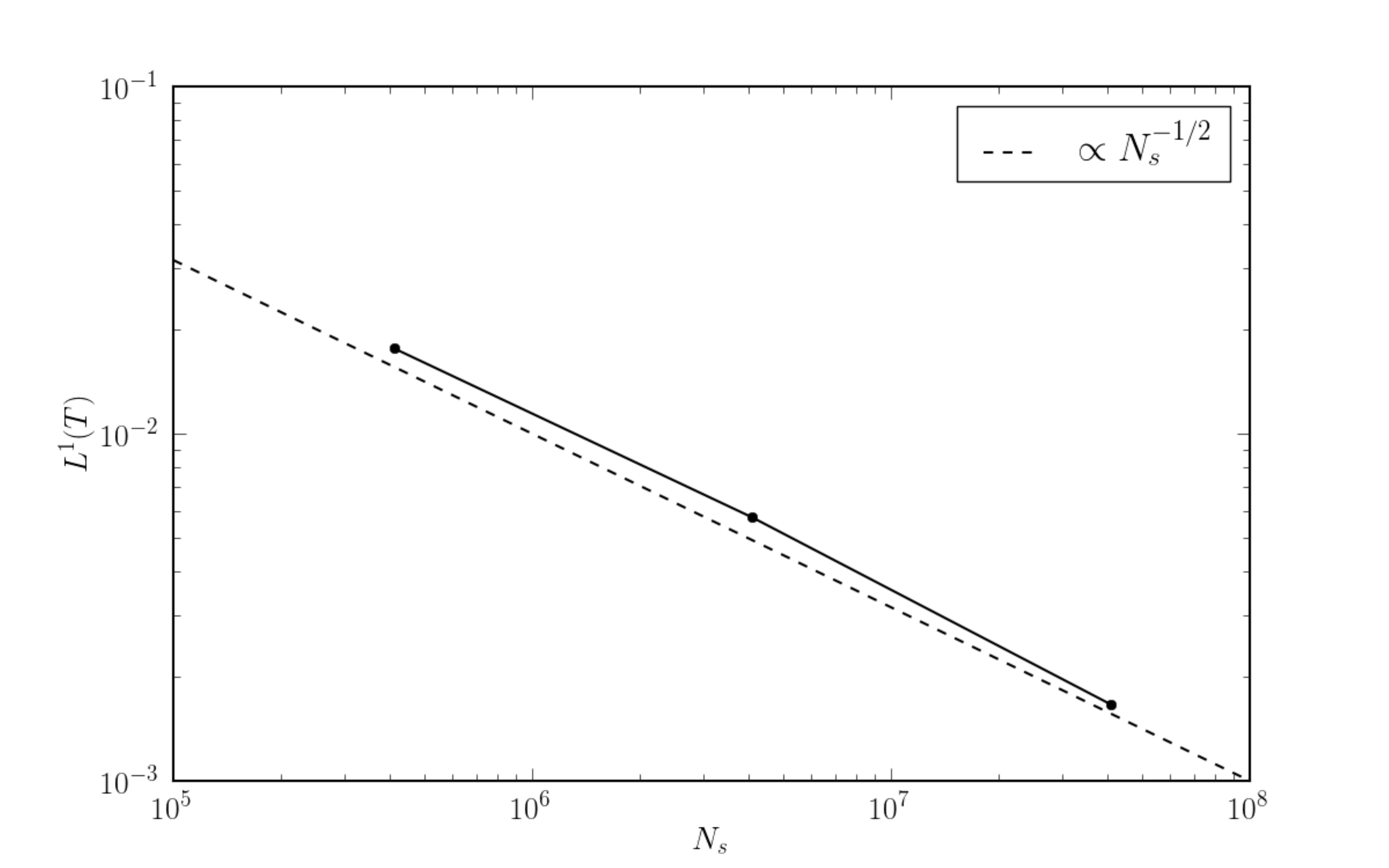}
\label{fig:transfermodeconvergence}
\end{figure}

We also examine convergence for $\tau = 1$ (with 100 grid zones); this result is shown in Figure \ref{fig:transfermodeconvergence}. Evidently the errors scale as $N_s^{1/2}$, as expected.

\subsection{Relativistic Radiation MHD Linear Modes}

We now also consider linear modes of the full equations of one-dimensional radiation magnetohydrodynamics; that is, we now include momentum exchange and magnetic fields. Acquiring even a linear solution to these equations with full transport is challenging, and so we resort to the approximate, relativistic Eddington closure scheme of \cite{farris2008} with gray opacity $\kappa$ from which to extract plane-wave solutions ${\bf P}$. Details of our calculation for a perturbation $\delta {\bf P}$ for $\delta \sim \exp(\omega t - ikx)$ about a thermal equilibrium ${\bf P}_0$ are given in Appendix \ref{linearmodes}. With a magnetic field $B^i = (B_0, B_0, 0)$ we follow the nonrelativistic treatment of \cite{jiang2012} by confining variation to a plane (suppressing Alfv\`en waves): ${\bf P} = (\rho, u, u^1, u^2, B^1, B^2, E, F^1, F^2) = (\rho_0 + \delta \rho, u_0 + \delta u, \delta u^1, \delta u^2, B_0, B_0 + \delta B^2, E_0 + \delta E, \delta F^1, \delta F^2)$. \bhl~is not designed to evolve perturbed equilibria; we focus on cases which accomodate both \bhl's numerical limitations as well as the discrepancy between the Eddington closure and \bhl's full transport (i.e.~we consider only many optical depths per wavelength). We study convergence of two specific cases with significant radiation pressure: a nonrelativistic radiation-modified slow MHD mode, and a relativistic radiation-modified fast MHD mode.  In this section, $c = k_B = 1$.  For all calculations, we use a box of length $L = 1$ with 128 grid zones, evolve to final time $t_f$, set the wavenumber of the perturbation $k = 2 \pi$, and normalize the $\delta {\bf P}$ to be $\lesssim 1 \%$ of the ${\bf P}_0$ for all ${\bf P}$.\footnote{The SageMath notebook used to evaluate these modes may be accessed via SageMathCloud at {\tt http://bit.ly/1CCi82y}} The ratio of radiation to gas pressure $\beta_r \equiv a_R P^3 / (3 \rho^4)$, and the optical depth per wavelength $\tau \equiv \kappa \rho L$. 

\subsubsection{Radiation-Modified Slow Mode}

\begin{figure}
  \caption{Convergence for the radiation-modified slow mode.}
  \centering
    \plotone{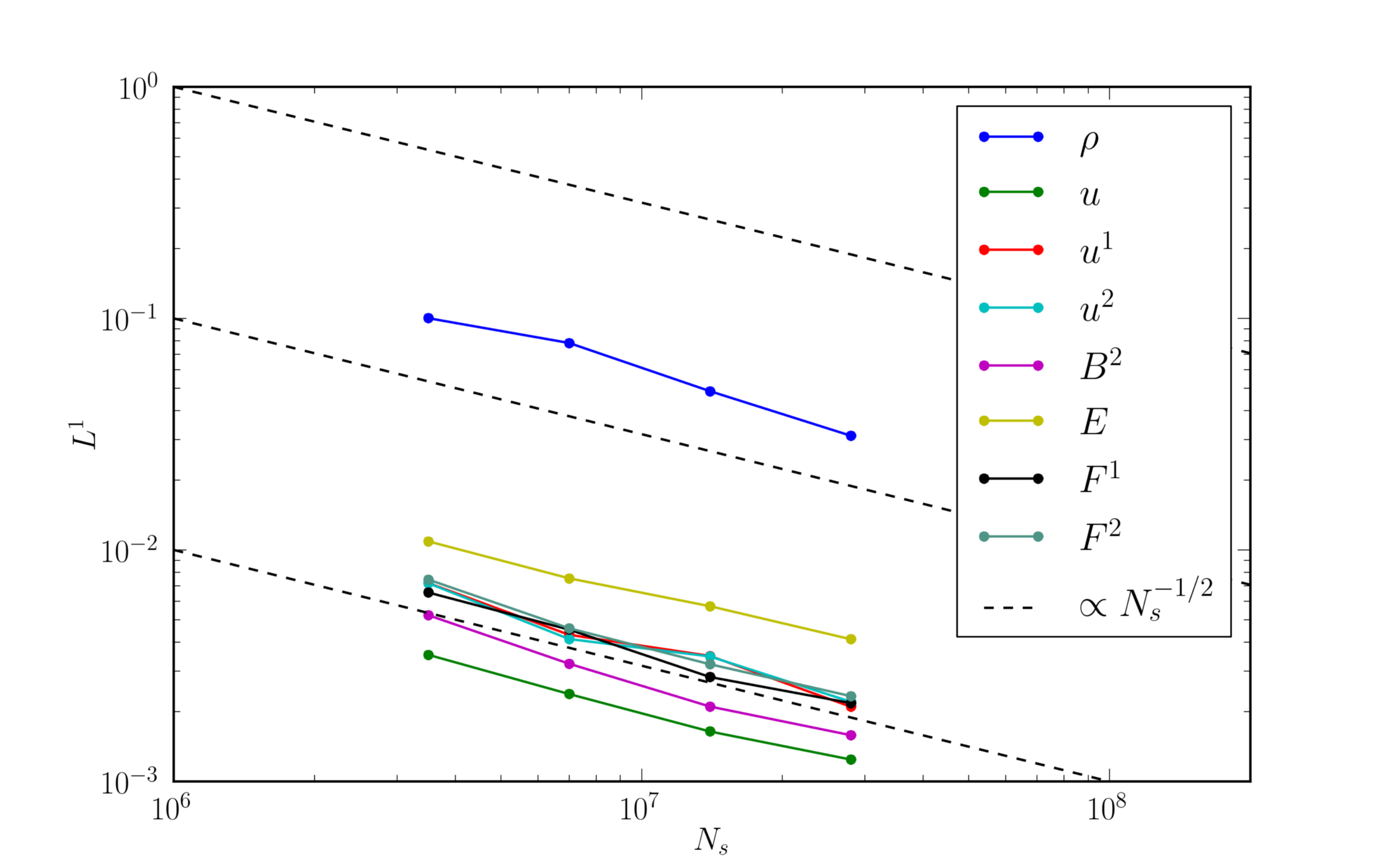}
\label{fig:slow_rmhd_mode_convergence}
\end{figure}

For a magnetized fluid in the presence of radiation, the MHD modes are damped in a similar fashion to the radiation hydrodynamic case. We first consider the radiation-modified slow mode solution. We set $\gamma = 5/3$, $\rho= 1$, $u = 0.01$, $B_0 = \sqrt{5/6}$, $\beta_r = 1$, and $\tau = 20$ and evolve the initial conditions in \bhl~to $t_f = 2.5$, nearly half an $e$-folding time. The eigenmode is given in Table \ref{table:radslowmode}. Expected convergence at $t_f$ in the average number of extant superphotons $N_s$ for Monte Carlo-dominated error is shown in Figure \ref{fig:slow_rmhd_mode_convergence}.

\begin{deluxetable}{cc}
\tablecolumns{5}
\tablewidth{0pt}
\tablecaption{Radiation-Modified Slow Mode}

\startdata
\hline
$\omega$ & $-0.155954250795 + 0.506371984839i$ \\
\hline
\hline
$\delta \rho$ & $0.992522043854$ \\
$\delta u$ & $0.0115437955397 +  0.00253571930238i$ \\
$\delta u^1$ & $-0.0799889439467-0.024635280384i $ \\
$\delta u^2$ & $-0.0804556011602-0.0252291311891i$ \\
$\delta B^2$ & $-0.00672014035309-0.00465692766557i$ \\
$\delta E$ & $0.0129233747759+0.0201394332108i$ \\
$\delta F^1$ & $0.00205715652365-0.00136719504591i$ \\
$\delta F^2$ & $-3.27455464963\times10^{-5}+5.02957595074\times10^{-5}i$ \\
\enddata
\label{table:radslowmode}
\end{deluxetable}

\subsubsection{Radiation-Modified Fast Mode}

\begin{figure}
  \caption{Convergence for the radiation-modified fast mode.}
  \centering
    \plotone{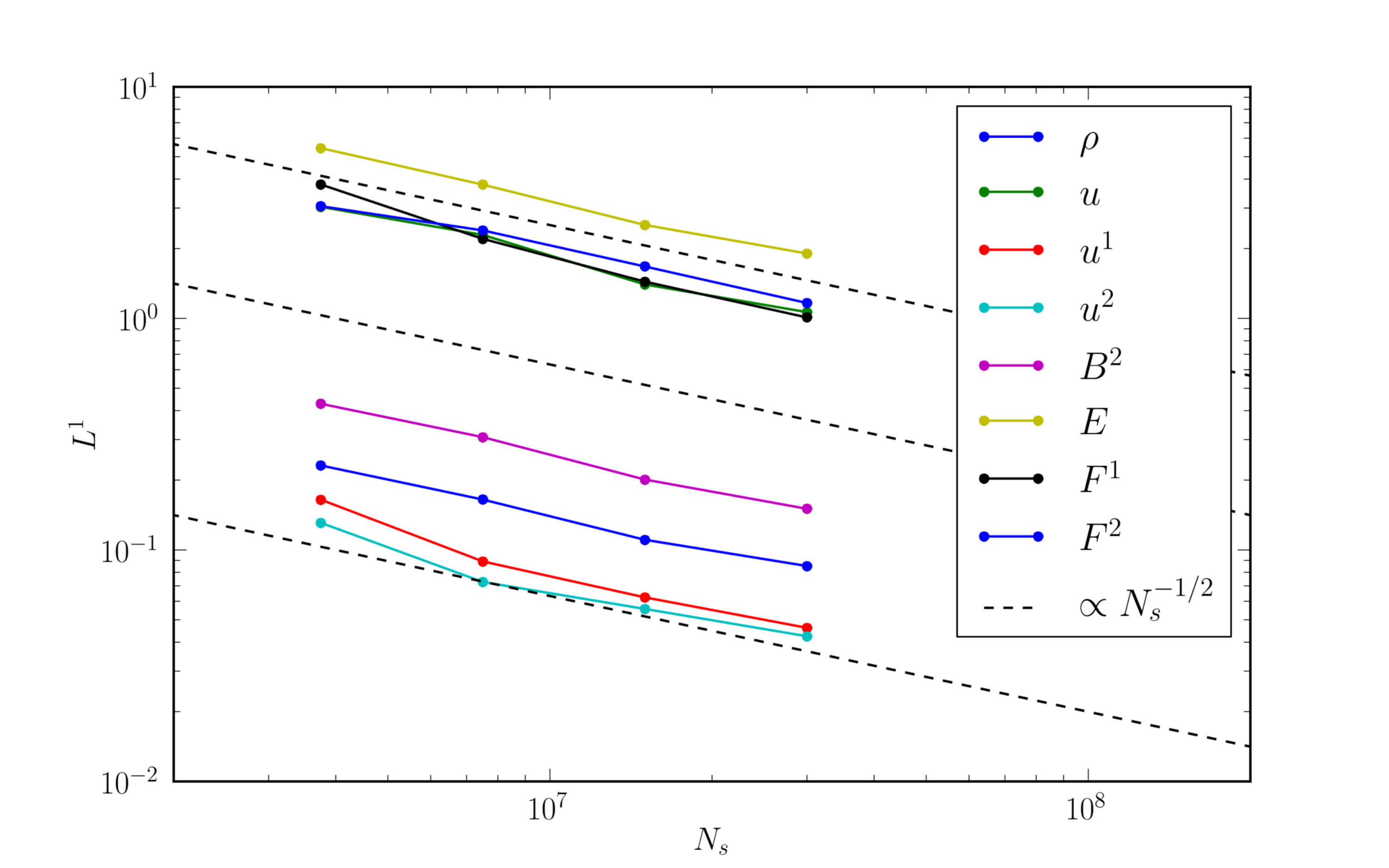}
\label{fig:fast_rmhd_mode_convergence}
\end{figure}

We now consider the radiation-modified fast mode solution, for a relativistic equilbrium. We set $\gamma = 4/3$, $\rho= 1$, $u = 10$, $B_0 = \sqrt{5/6}$, $\beta_r = 1$, and $\tau = 20$ and evolve the initial conditions in \bhl~to $t_f = 1.7$, approximately a wave period. Note that radiation damping is not significant during this time. The eigenmode is given in Table \ref{table:radfastmode}. Expected convergence at $t_f$ in the average number of extant superphotons $N_s$ for Monte Carlo-dominated error is shown in Figure \ref{fig:fast_rmhd_mode_convergence}.

\begin{deluxetable}{cc}
\tablecolumns{5}
\tablewidth{0pt}
\tablecaption{Radiation-Modified Fast Mode}

\startdata
\hline
$\omega$ & $-0.000695187092855 - 3.6761842859i$ \\
\hline
\hline
$\delta \rho$ & $0.0528655837266-0.000203781373769i$ \\
$\delta u$ & $0.704834313006 -0.00203965222393i$ \\
$\delta u^1$ & $0.0309307265333-0.000125078175647i $ \\
$\delta u^2$ & $-0.00191512771962+ 0.000183786243559i$ \\
$\delta B^2$ & $ 0.0512475714061 -0.000472212042911i$ \\
$\delta E$ & $0.704838422737$ \\
$\delta F^1$ & $-4.41864621469\times10^{-5}+0.00198608271502i$ \\
$\delta F^2$ & $0.000397994468532-0.00462049532989i$ \\
\enddata
\label{table:radfastmode}
\end{deluxetable}

\subsection{Su-Olson Problem}

\cite{suolson1996} found a solution in terms of integrals to the coupled energy balance and radiative transport equations in the diffusion approximation for a semi-infinite slab of static, initially cold fluid (with heat capacity $c_v =  \mathscr{C} T^3$) with gray absorption coefficient $\alpha$, and a Marshak (isotropic incident radiation) condition at the left boundary with incident flux $F$. The solution is given in terms of dimensionless gas and radiation energy densities $u$ and $v$, respectively, versus the dimensionless spatial coordinate $x = \sqrt{3}\alpha z$ and dimensionless time coordinate $t = 4 a_R c \alpha t'/\mathscr{C}$, where $a_R$ is the radiation constant. $u$ and $v$ are defined as:
\begin{equation}
u(x, t) \equiv \left(\frac{c}{4}\right)\left(\frac{E(x, t)}{F}\right),
\end{equation}
\begin{equation}
v(x, t) \equiv \left(\frac{c}{4}\right)\left(\frac{a_R T^4(x,t)}{F}\right),
\end{equation}
where $E$ is the radiation energy density.

\begin{figure}
  \caption{Numerical gas and radiation energy densities versus analytic gas and radiation energy densities. The \bhl~calculation and Su-Olson results show excellent correspondence at this late time.}
  \centering
    \plotone{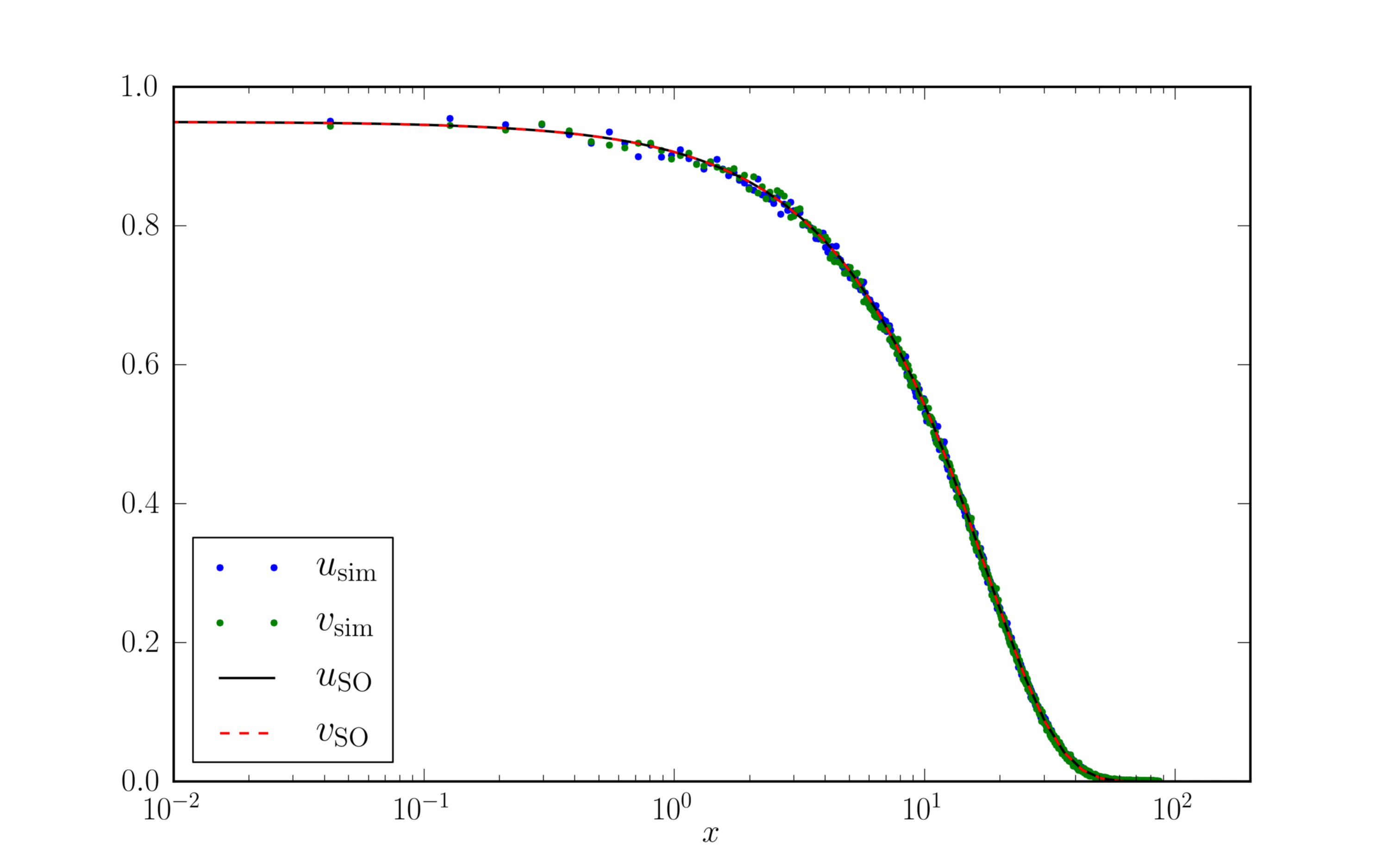}
\label{fig:suolson}
\end{figure}

In replicating this solution with \bhl, we adopt parameters such that the solution remains optically thick, without being so optically thick across a grid zone that our Monte Carlo transport scheme fails. We use 1024 grid zones on $x = [0, 50\sqrt{3}]$ and evolve the system to $t = 500$ (when the solution is approximately in equilibrium). Results are shown in Figure \ref{fig:suolson}.

\subsection{Radiative Shocks}
\label{sec:radiativeshocks}

We now turn to dynamical tests: here, 1D radiative shocks.  We will use nearly the same suite of relativistic, radiative shocks considered by \cite{farris2008} in testing their GRRMHD code, which was based on a nonequilibrium Eddington closure.  Because we solve the full transfer equation, we expect disagreement on scales of order the photon mean free path.

The Farris et al. tests assume a grey opacity $\kappa$ and are set in Minkowski space.  The solutions are described by $\rho_0$, the $x$ component of the fluid four velocity $u^x$, the gas pressure $P$, and the comoving-frame radiation energy density $E$ and $x$-component of the radiation flux four-vector $F^x$.  The latter vanishes far from the shock.  We consider only the shock-frame (unboosted) version of the tests, initialized as shock tubes.   All tests are purely hydrodynamic: the magnetic field plays no role.

Our tests are identical to those given in \cite{farris2008} except that we modify case (4), which has radiation pressure a factor of 10 larger than gas pressure upstream from the shock.  Due to this large radiation pressure, \bhl~cannot integrate this case stably with the numerical resources available to us.  Instead, we set the upstream radiation pressure equal to the gas pressure, and call this case (4a).  The shock parameters are listed in Table \ref{table:farrisparams}.  Units are such that $c=1$; $a_R$ (equivalently, $\hbar$) is determined by enforcing thermal equilibrium $E = a_R (P/\rho_0)^4$ far from the shock.

\begin{deluxetable}{cclll}
\tablecolumns{5}
\tablewidth{0pt}
\tablecaption{Parameters for Farris shocks}
\tablehead{\colhead{Case} & \colhead{$\gamma$} & \colhead{$\kappa$} & \colhead{Left State} & \colhead{Right State} } 

\startdata
1  & 5/3 & 0.4 & $\rho_0 = 1.0$ & $\rho_0 = 2.4$\\
 & & & $P = 3.0\times10^{-5}$& $P = 1.61\times10^{-4}$\\
 & & & $u^x = 0.015$ & $u^x = 6.25\times10^{-3}$ \\
 & & & $E = 1.0\times10^{-8}$ & $E = 2.51\times10^{-7}$\\ \hline
2 & 5/3 & 0.2 & $\rho_0 = 1.0$ & $\rho_0 = 3.11$\\ 
 & & & $P = 4.0\times10^{-3}$& $P = 0.04512$\\
 & & & $u^x = 0.25$ & $u^x = 0.0804$ \\
 & & & $E = 2.0\times10^{-5}$ & $E = 3.46\times10^{-3}$\\ \hline
3 & 2 & 0.3 & $\rho_0 = 1.0$ & $\rho_0 = 8.0$\\ 
 & & & $P = 60.0$& $P = 2.34\times10^{3}$\\
 & & & $u^x = 10.0$ & $u^x = 1.25$ \\
 & & & $E = 2.0$ & $E = 1.14\times10^{3}$\\ \hline
4a & 5/3 & 0.4 & $\rho_0 = 1.0$ & $\rho_0 = 1.165$\\
 & & & $P = 0.1$& $P = 0.1233 $\\
 & & & $u^x = 0.5$ & $u^x = 0.4292$ \\
 & & & $E = 0.3$ & $E = 0.3763$\\ \hline
\enddata
\label{table:farrisparams}
\end{deluxetable}

\begin{figure}
  \caption{Gas and radiation variables for radiative shock Case 1. The result from \bhl~is shown as a red solid line, and the analytic solution is shown as a dashed line.}
  \centering
    \plotone{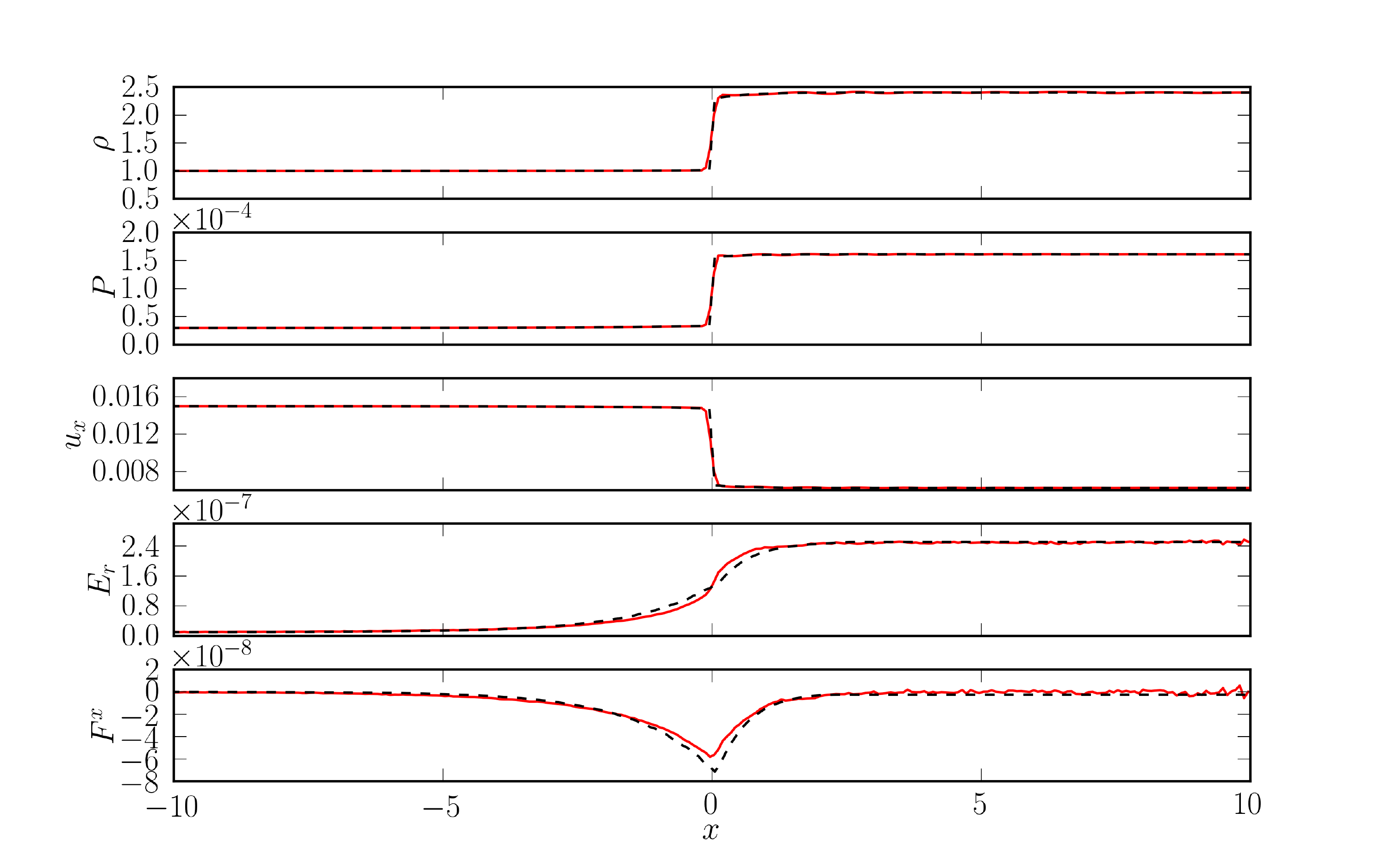}
\label{fig:farris1}
\end{figure}

Case 1 is a nonrelativistic strong shock with gas pressure much greater
than radiation pressure, and consequently the fluid variable profiles
resemble a nonrelativistic shock. \bhl~output and the analytic solution
are shown in Figure \ref{fig:farris1}. Correspondence is good except for
small deviations in the radiation variables near the shock interface, as
expected for our full transfer solution.

\begin{figure}
  \caption{Gas and radiation variables for radiative shock Case 2. The result from \bhl~is shown as a red solid line, and the analytic solution is shown as a dashed line.}
  \centering
    \plotone{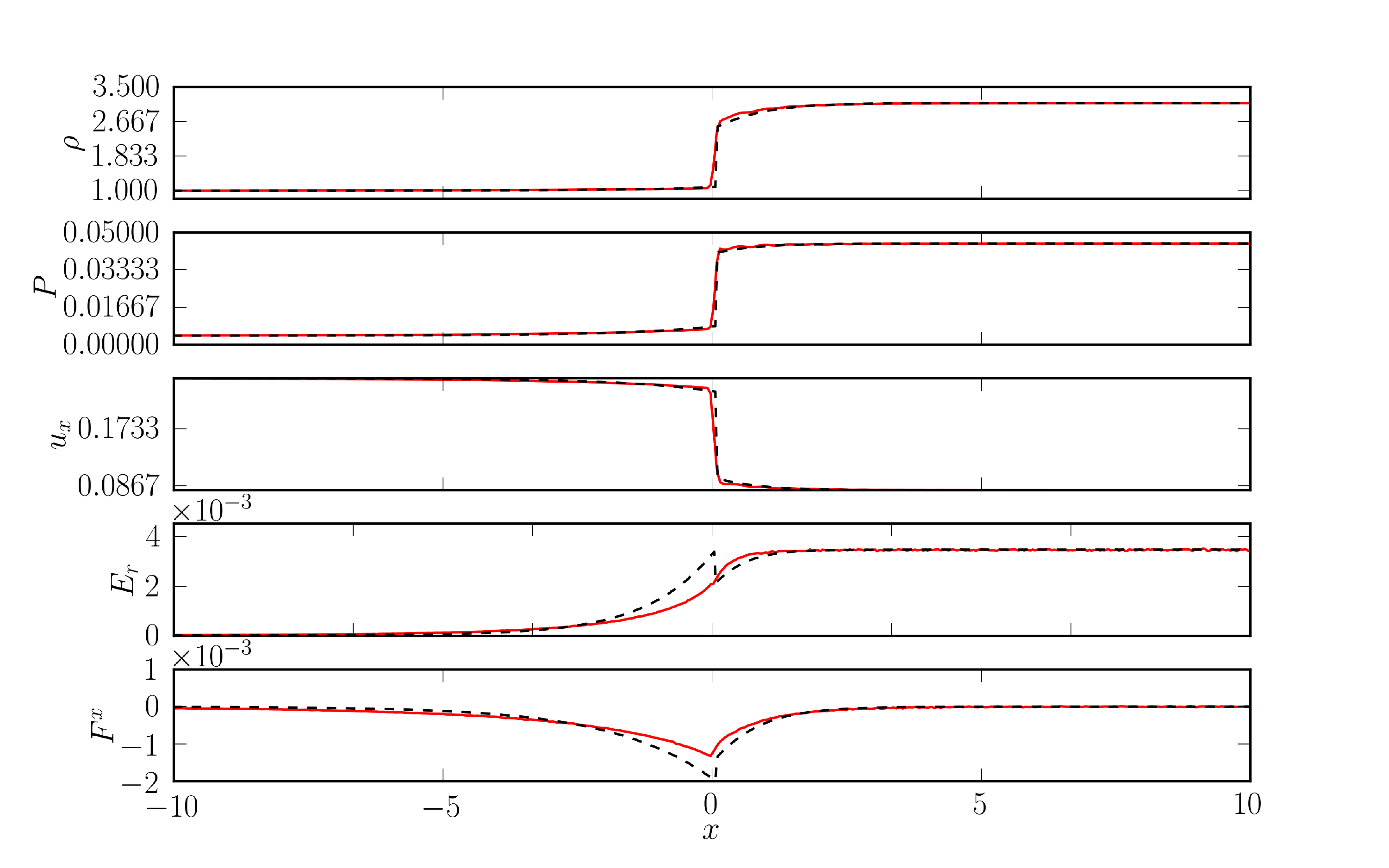}
\label{fig:farris2}
\end{figure}

Case 2 is a mildly relativistic shock with somewhat larger radiation pressure than in Case 1. \bhl~output and the analytic solution are shown in Figure \ref{fig:farris2}. The profiles of $E$ and $F^x$ show qualitative differences between the \bhl~result and the analytic solution: a discontinuity in the analytic solution does not appear for the case of full transfer.  This is unsurprising given the approximate nature of the Farris solutions. Note that for this case the approximate Eddington solution contains an unphysical discontinuity in the coordinate frame radiation energy density.

\begin{figure}
  \caption{Gas and radiation variables for radiative shock Case 3. The result from \bhl~is shown as a red solid line, and the analytic solution is shown as a dashed line.}
  \centering
    \plotone{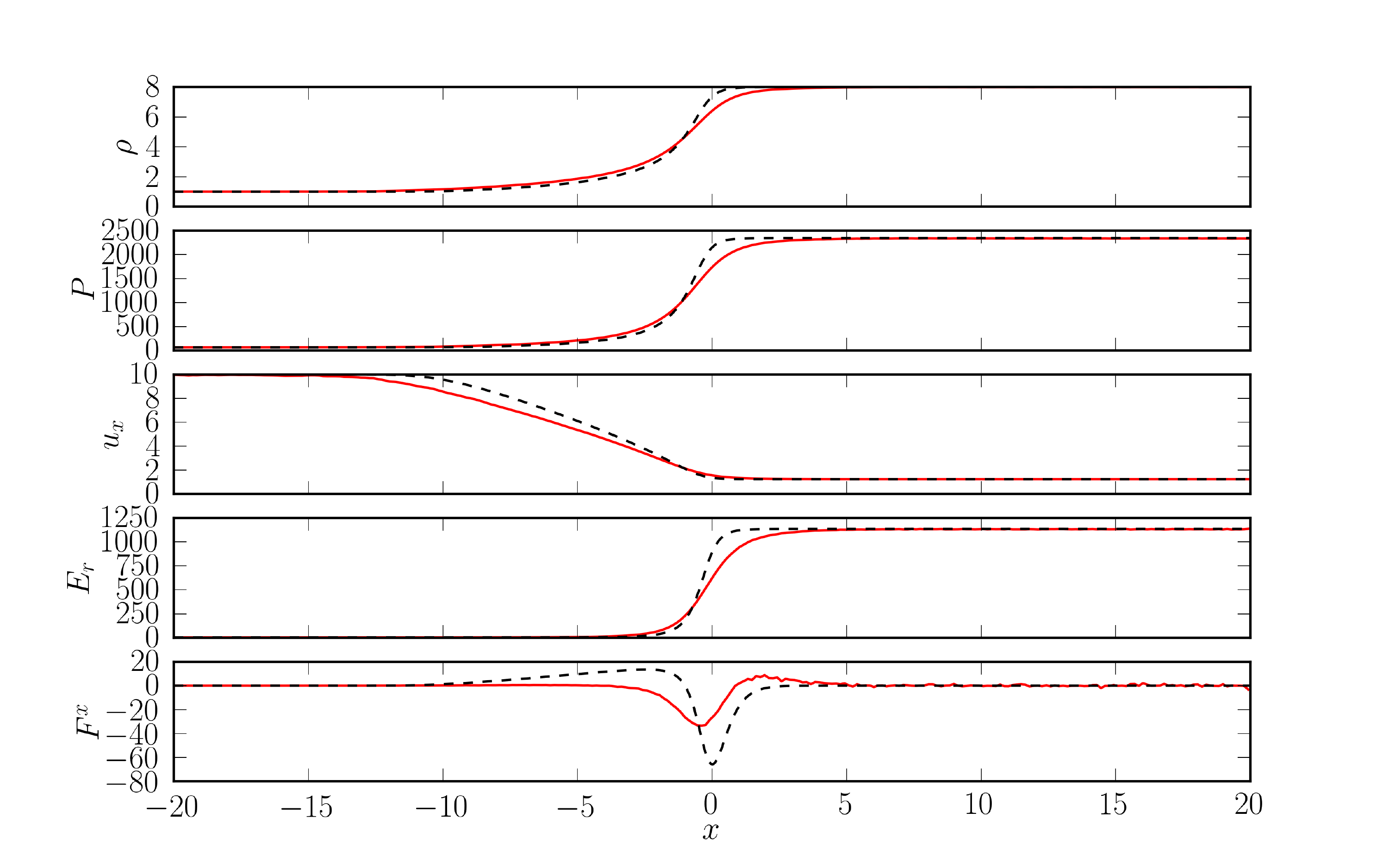}
\label{fig:farris3}
\end{figure}

\begin{figure}
  \caption{Self-convergence of all variables in radiative shock Case 3. $N_s$ is directly proportional to the number of extant photons in the simulation. Dashed lines correspond to the $N_s^{-1/2}$ trend expected for Monte Carlo integration in the absence of resolution errors in the hydrodynamics solver.}
  \centering
    \plotone{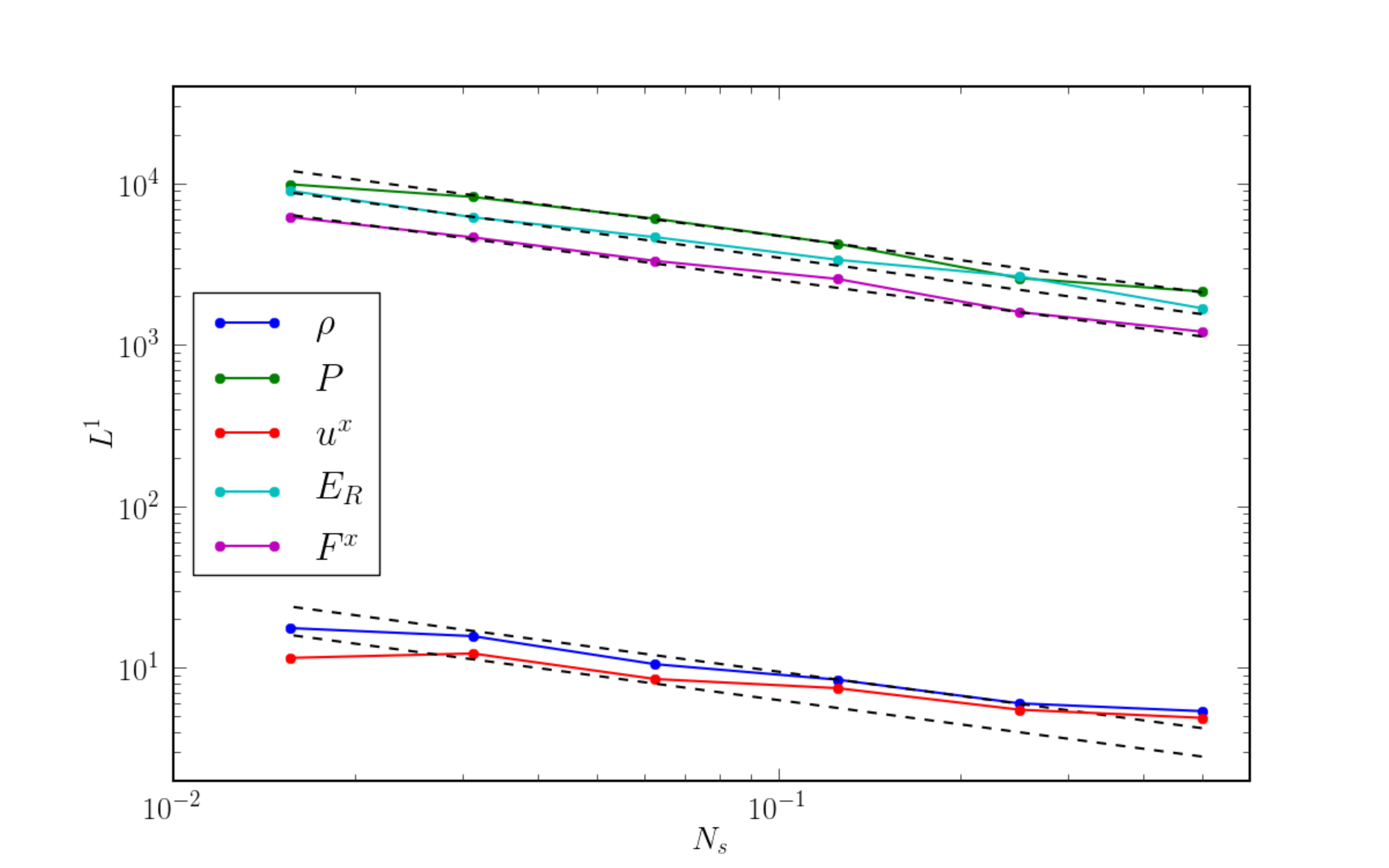}
\label{fig:farris3convergence}
\end{figure}

Case 3 is a highly relativistic shock with dynamically important radiation field. \bhl~output and the analytic solution are shown in Figure \ref{fig:farris3}. We find significant differences within a few photon mean free paths of the shock, particularly for the radiation flux. Figure \ref{fig:farris3convergence} shows the expected $N^{-1/2}$ self-convergence in the radiation variables. A similar self-convergence trend also appears in the fluid variables until grid resolution becomes the dominant source of error. 

\begin{figure}
  \caption{Gas and radiation variables for radiative shock Case 4a. The result from \bhl~is shown as a red solid line, and the analytic solution is shown as a dashed line.}
  \centering
    \plotone{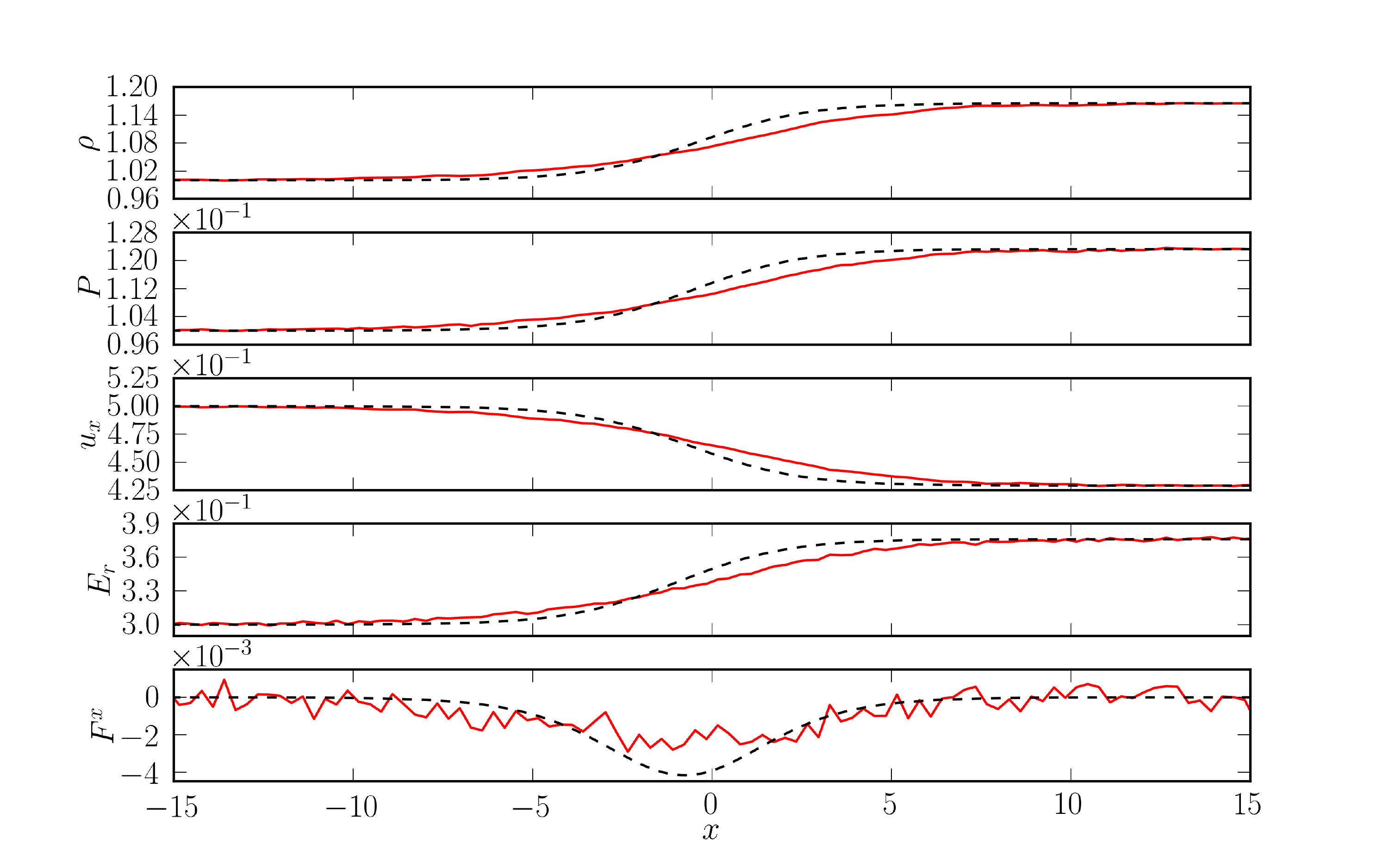}
\label{fig:farris4a}
\end{figure}

Case 4a is a modestly relativistic wave with upstream radiation and gas pressure nearly equal. \bhl~output and the analytic solution are shown in Figure \ref{fig:farris4a}. We find good agreement in all variables, despite the relatively strong radiation field. Note also that even with a large number of samples it is difficult to suppress noise in $F^x$ when it is much smaller than $E$.

\subsection{Black Hole Atmosphere}
\label{sec:atmosphereproblem}

Next we turn to a general relativistic equilibrium test.  Consider a Schwarzschild black hole surrounded by a static atmosphere.  The atmosphere is bounded by static, concentric spherical shells at $r_i > 2 GM/c^2$ and $r_o > r_i$, and is in radiative equilibrium.  The shells are reflecting boundaries and exchange no heat with the atmosphere, which has a grey opacity $\kappa$ and adiabatic index $\gamma = 5/3$.  

In the Newtonian limit radiative conduction would drive the atmosphere toward $T = \mathrm{const}.$  In a relativistic atmosphere the gravitational field causes the atmosphere to come into a different equilibrium in which redshifted temperature, $T_\infty \equiv T\sqrt{-g_{00}}$, is constant. 

In the Eddington approximation the atmospheric structure is determined by the conditions $T_\infty = \mathrm{const}$, and mechanical equilibrium, $T^{r r}_{~~;r} = 0$. Note that any radiation source terms vanish in thermal equilibrium; in fact, the Eddington approximation solution turns out to be an exact solution to the transfer equation, as we show in Appendix \ref{apB}.  

\begin{figure}
  \caption{Rest-mass density and both gas and radiation temperatures for the static atmosphere test at $t = 150M$. Dashed lines represent the analytic solutions.}
  \centering
    \plotone{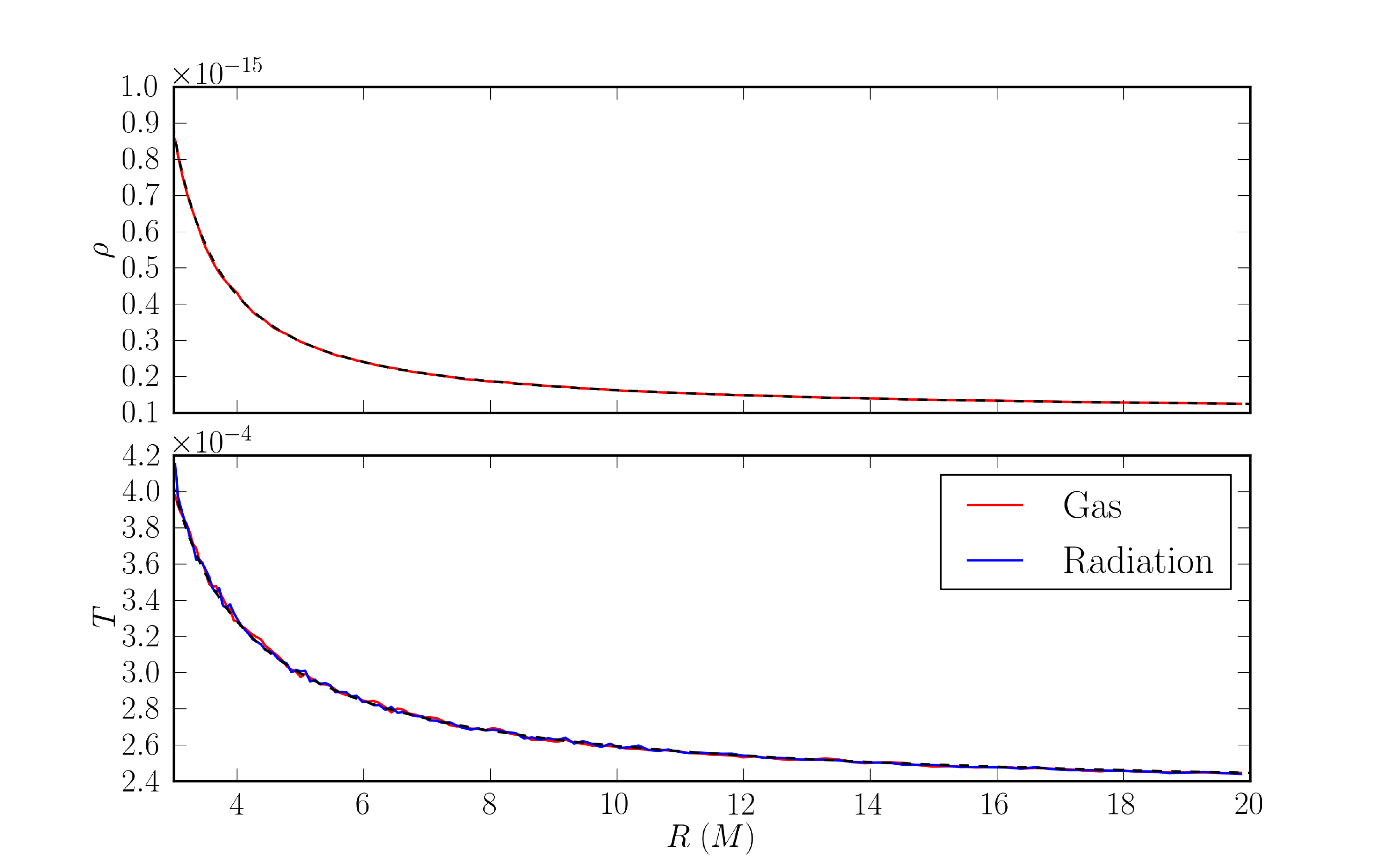}
\label{fig:ksatmos}
\end{figure}

\begin{figure}
  \caption{Spectrum of superphotons at $r \approx r_i$ for the radiating atmosphere test, showing good correspondence to the expected Planck spectrum.}
  \centering
  \plotone{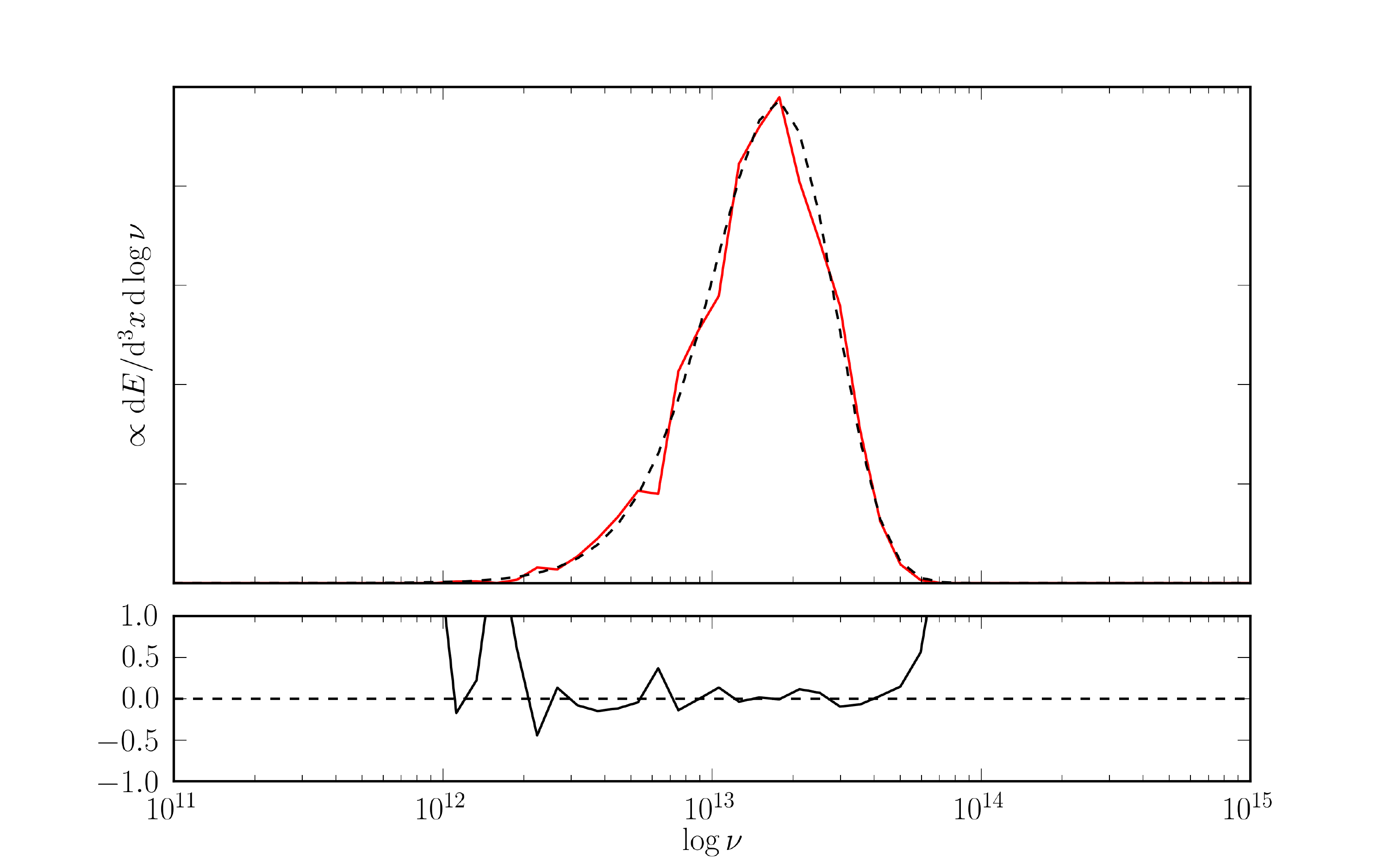}
\label{fig:ksatmos_spectrum}
\end{figure}

Once the inner and outer radii are specified, there are three dimensionless parameters that describe the solution (although their interpretation is purely Newtonian): the ratio of the atmospheric scale height at the inner boundary to the local radius $h_i \equiv \gamma k_B T_i r_i / (\mu m_p G M)$ (where $\mu$ is the mean molecular weight), the ratio of radiation pressure to gas pressure at the inner boundary $\beta_r \equiv \mu m_p a_R T_i^3 / (3 \rho_i k_B)$, and the characteristic optical depth $\tau  \equiv \kappa \rho_i (r_o - r_i)$. 

We set $r_i = 3GM/c^2$, $r_o = 20GM/c^2$.  We set $h \approx 2.66$, $\beta_r \approx 0.23$, and $\tau \approx 5.0$. We use a 1D domain with 128 zones; the errors are dominated by Monte Carlo noise. The solution is shown in Figure \ref{fig:ksatmos}. We also find the comoving spectrum at fiducial radius $r \approx r_i$ to match our expectation for thermalized radiation (with free vertical scale), as shown in Figure \ref{fig:ksatmos_spectrum}.  Evidently the redshifted temperature is indeed constant.

\section{Applications}

\subsection{Radiating Bondi Accretion}

Here we report a preliminary investigation of spherically symmetric accretion onto a Schwarzschild black hole with radiative coupling.  The fully dynamical problem has been studied analytically (although not with a full transport solution, frequency dependence, or magnetic fields) by \cite{vitello1984}, \cite{park1990}, and \cite{nobili1991}. Frequency-integrated numerical studies have also been performed in \cite{fragile2012}, \cite{roedig2012}, and \cite{mckinney2013} over many $GM/c^2$.  Here we obtain a full transport solution with frequency dependence and magnetic fields.  We include synchrotron emission, synchrotron absorption, and Compton scattering, with associated heating and cooling.

We set $M = 6.6\times10^9 M_{\odot}$ and describe the spacetime with modified Kerr-Schild (MKS) coordinates (Section \ref{sec:miscellaneous}). All simulations are performed in 1D, with $r \in [1.5M,50M]$ for 64 zones. As initial conditions for the fluid, we adopt the nonradiative Bondi solution of \cite{hawley1984} (which we hold constant at the outer boundary), except we set $\gamma = 13/9$ and place the sonic point at $r=200M$. We vary the accretion rate by varying the density (or equivalently the mass unit $\munit$) of the flow.

The magnetic field is initialized as a radial, monopolar field, which has no effect on the fluid motion.  We set $B^1 = \alpha/r^3$, where $\alpha$ is chosen such that $\beta \equiv 2 P/b^2 \approx 130$ at $r = 2M$ in the initial (GRMHD) conditions. No radiation is present initially.  The radiation is allowed to flow freely out of the computational domain at the radial boundaries, with no inflow of radiation.

The luminosity $L(r)$, evaluated once the flow has settled and become almost time-independent, is
\begin{equation}
L(r) \equiv \int \sqrt{-g}\, \mathrm{d}x^2 \mathrm{d}x^3 R^1_{~0}.
\label{eqn:luminosity}
\end{equation}
We average $L(r)$ over radial shells between $r=30GM/c^2$ and $r=50GM/c^2$ to obtain an average luminosity $L$.  This is equivalent to time-averaging at a single radius.   

\begin{figure}
  \caption{Fluid and radiation profiles for the radiating Bondi problem. Case 1 is shown in purple, Case 2 in teal, Case 3 in red, Case 4 in green, and Case 5 in blue. The dashed line shows Case 5 without Compton scattering.}
  \centering
    \plotone{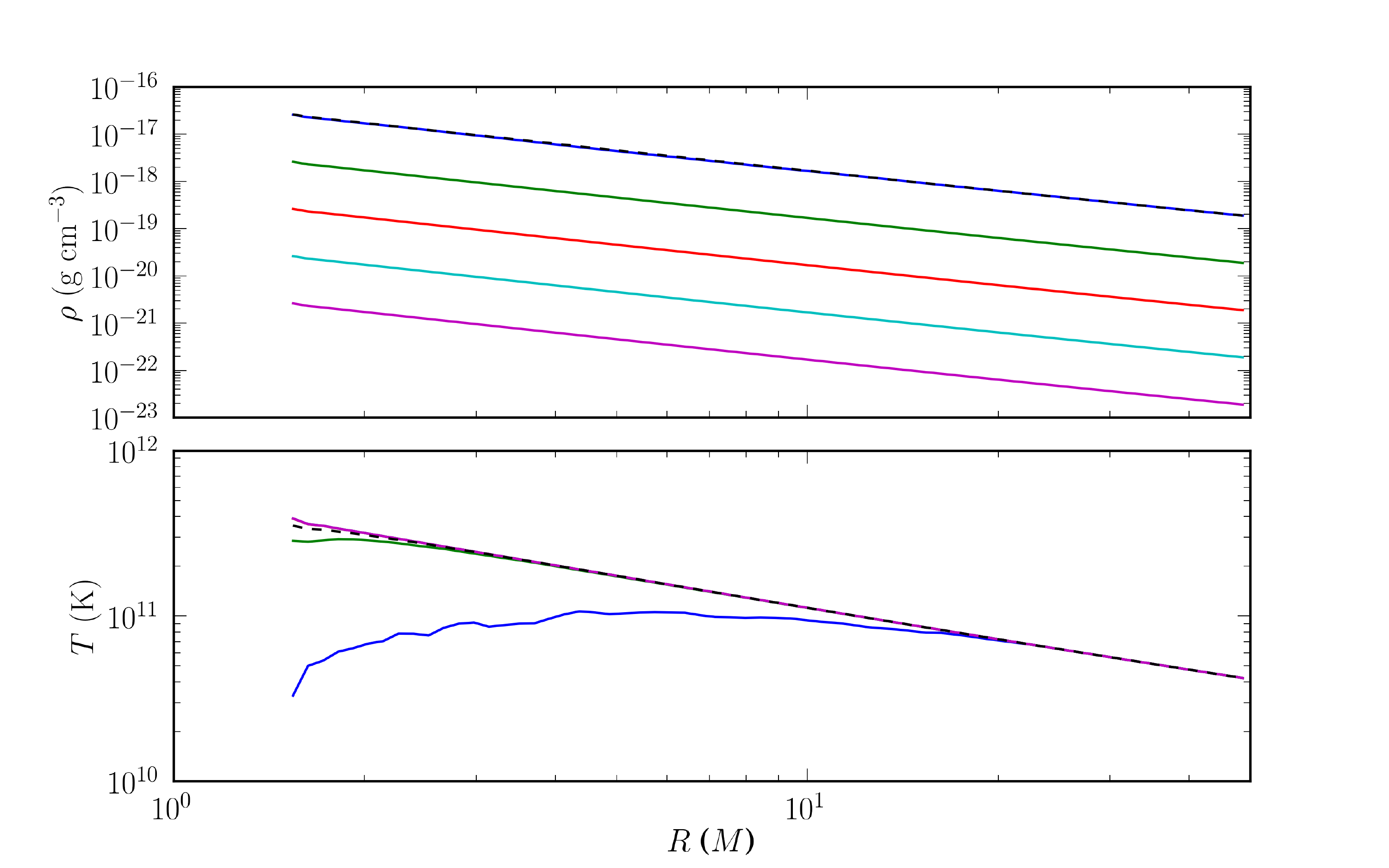}
\label{fig:synchbondi}
\end{figure}

\begin{figure}
  \caption{Efficiency of accretion for the radiating Bondi problem.}
  \centering
    \plotone{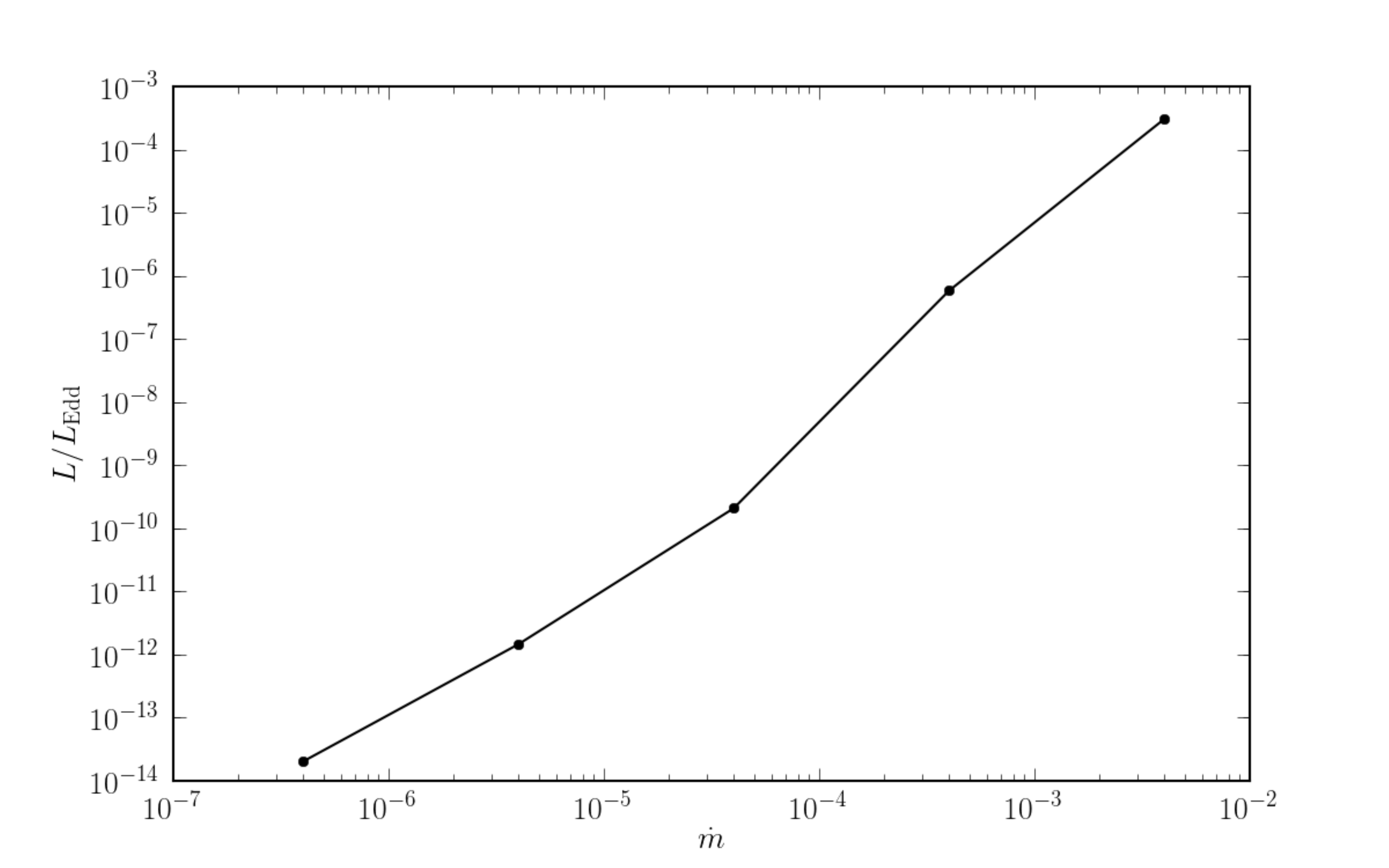}
\label{fig:bondiefficiency}
\end{figure}

We perform five calculations at different accretion rates. We evolve each system until the luminosity becomes stable.  At high accretion rate this can take as long as $400GM/c^3$. The characteristic number density for each simulation, mass accretion rate
\begin{equation}
\dot{M} \equiv -\int \sqrt{-g} \,\mathrm{d}x^2 \mathrm{d}x^3 \rho u^1
\label{eqn:mdot}
\end{equation} 
(evaluated at the event horizon), and L are given in Table \ref{table:bondiparams}. The resulting profiles of these five cases, along with that of Case 5 with Compton scattering disabled, are shown in Figure \ref{fig:synchbondi}. The lack of cooling in the pure synchrotron case indicates that Compton cooling dominates over synchrotron cooling for the highest accretion rate model. The relationship between luminosity and accretion rate when Compton scattering is active is shown in Figure \ref{fig:bondiefficiency}.

\begin{figure}
  \caption{Self-convergence for the radiating Bondi problem near the Eddington limit. Dashed lines show convergence $\propto N_s^{-1/2}$.}
  \centering
    \plotone{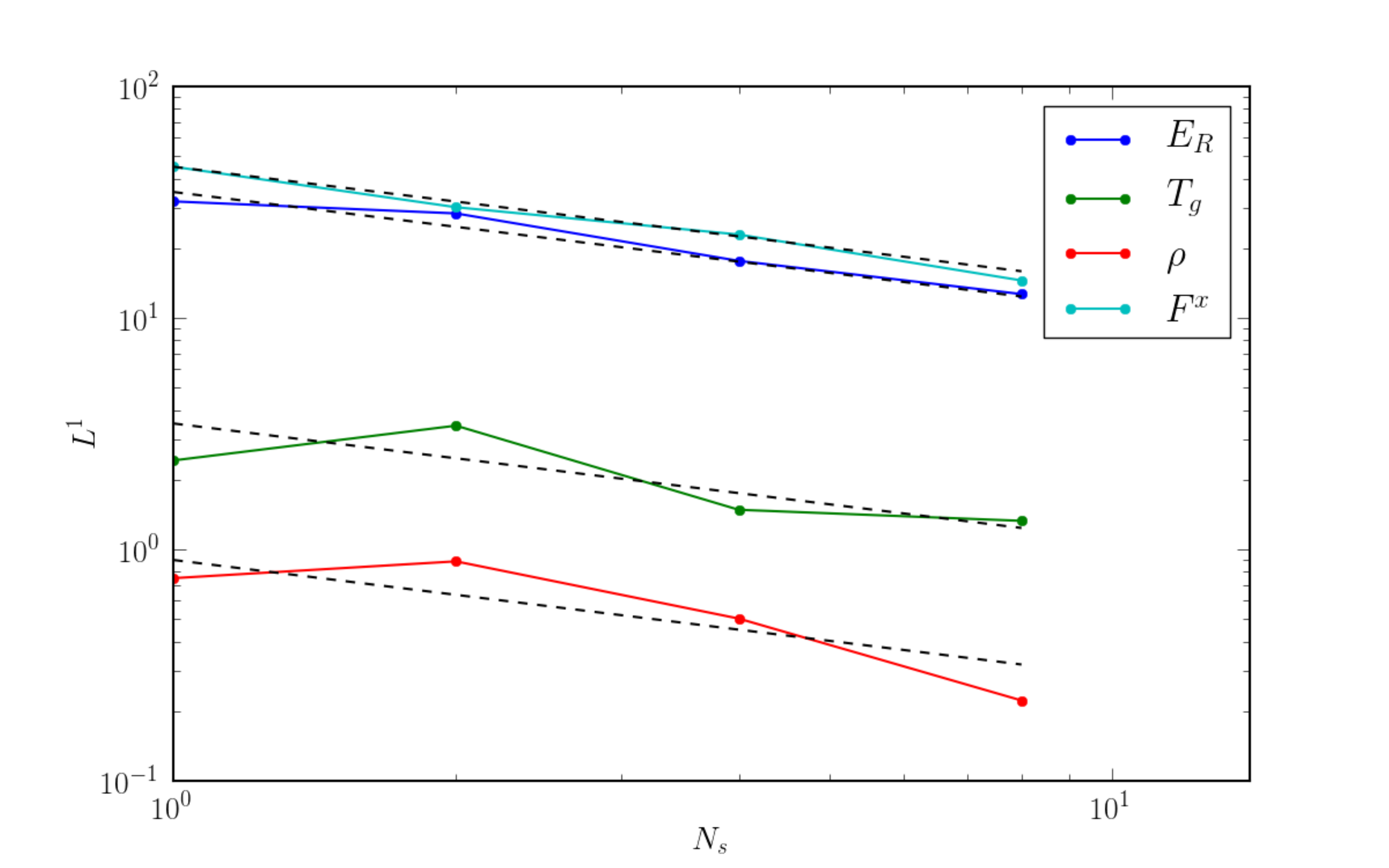}
\label{fig:bondiconvergence}
\end{figure}

We have checked self-convergence of a solution with $n = 1.51\times10^{10}$ cm$^{-3}$ (between Cases 4 and 5 in Table  \ref{table:bondiparams}) in steady state at $t=200M$. The expected convergence behavior for Monte Carlo-dominated error is shown in Figure \ref{fig:bondiconvergence}.

Our models show that Compton cooling is important for Bondi accretion near the Eddington rate, and that -- for our assumed field configuration -- synchrotron cooling is comparatively unimportant.   Although here we find appreciable cooling only close to the Eddington rate, we expect that for near-Keplerian accretion flows Compton scattering will become dynamically important at lower accretion rates, as individual fluid elements will have more time to cool (their radial velocity is lower) before accreting onto the black hole.

\begin{deluxetable}{cccc}
\tablecolumns{4}
\tablewidth{0pt}
\tablecaption{Parameters for radiating Bondi accretion}
\tablehead{\colhead{Case} & \colhead{$n$} & \colhead{$\dot{m}$} & \colhead{$L$} \\ 
\colhead{} & \colhead{cm$^{-3}$} & \colhead{} &\colhead{$L_{\mathrm{Edd}}$} } 

\startdata
1  &  $3.0\times10^6$  & $4.01\times10^{-7}$ & $2.03\times10^{-14} $ \\
2  &  $3.0\times10^7$  & $4.01\times10^{-6}$ & $1.46\times10^{-12} $ \\ 
3  &  $3.0\times10^8$  & $4.01\times10^{-5}$ & $2.08\times10^{-10} $ \\
4  &  $3.0\times10^{9}$  & $4.01\times10^{-4}$ & $5.81\times10^{-7}$ \\ 
5  &  $3.0\times10^{10}$  & $4.01\times10^{-3}$ & $3.05\times10^{-4}$
\enddata
\label{table:bondiparams}

\end{deluxetable}

\subsection{Axisymmetric Radiating Kerr Black Hole Accretion}
\label{sec:torus}

As another preliminary application of \bhl~we consider the effect of radiation on an intermediate accretion rate black hole accretion flow.  Recall that for systems with $L \sim 10^{-9} \Ledd$ (like Sgr A*) radiation will have little effect since the cooling timescale is long compared to the accretion timescale.  This is the classical RIAF regime.  For systems with $L \sim 10^{-5} \Ledd$ (like M87) the effect of radiation depends on temperature and the distribution function of the electrons, but in certain circumstances radiation interactions -- especially Compton cooling -- can cool the flow on timescales comparable to or shorter than the accretion timescale.

Here we consider an axisymmetric accretion flow with parameters inspired by M87.  We set black hole mass $M = 6.6\times10^9 M_{\odot}$ following \cite{gebhardt2011} and dimensionless spin $a_* = 0.9375$, and we adjust $\munit$ so that $\dot{m} \approx 6.3\times10^{-6}$ \citep{kuo2014}.  Although the distribution functions for ions and electrons in M87-like systems probably contains multiple components, we adopt thermal distributions for both and set the ion and electron temperatures $T_i = 3T_e$. $T_i/T_e$ may strongly affect the dynamics of this system; essentially, it controls the cooling rate.  We set the adiabatic index $\gamma=13/9$, appropriate for ionized hydrogen when $kT_i \ll m_p c^2$ and $kT_e \gg m_e c^2$.  We include synchrotron emission for a relativistic, thermal distribution of electrons (emissivity given by Equation 72 of \citealt{leung2011}), thermal absorption, and Compton scattering.  Bremsstrahlung is neglected, as it is a small correction to the emissivity within $\sim 10^3 M$ of the black hole for such intermediate accretion rate systems (e.g. \citealt{narayanyi1999}). 

The initial conditions are a Fishbone-Moncrief torus (\citealt{fishbonemoncrief1976}) with an inner radius at $6GM/c^2$ and pressure maximum at $12GM/c^2$. We extend this configuration by adding a weak poloidal magnetic field that follows isodensity contours, using the same procedure as in \cite{mckinney2004}, but with the vector potential modified by a $\cos \theta$ factor to produce a two-loop configuration.  The field strength is normalized so that $\beta$ has a global minimum value of 100.  Small perturbations are applied to the internal energy to efficiently initiate the magnetorotational instability. No radiation is present in the initial conditions.

We use the modified Kerr-Schild coordinates (see \S \ref{sec:miscellaneous}), with $r_0 = 0$ and $h_s = 0.3$.  The grid runs from $r = 0.98(1+\sqrt{1-a_*^2})GM/c^2$ to $r = 40GM/c^2$, and from $\theta = 0$ to $\theta = \pi$.  We impose outflow boundary conditions on both the fluid and radiation at the inner and outer radial boundaries, and reflecting polar boundary conditions. In this instance we adopt the piecewise parabolic method for reconstruction. We set the CFL number to $0.7$, and evolve the system until $t = 2000M$.

To accurately sample cooling due to scattering events, we bias the scattering via the method given in Section \ref{sec:scattering} such that $b_s \propto \Theta_e^2$. To determine the absolute number of superphotons in steady state, we require only that the bolometric light curve be satisfactorily resolved (here, $\sim 1.1\times10^6$ superphotons at any one time). In the future we will more carefully consider the resolution requirement for these models.

\begin{figure}
  \caption{Torus temperature $\Theta_e$ and comoving radiation energy density $R^{\mu\nu}u_{\mu}u_{\nu}$ at $t=1500M$.}
  \centering
    \plotone{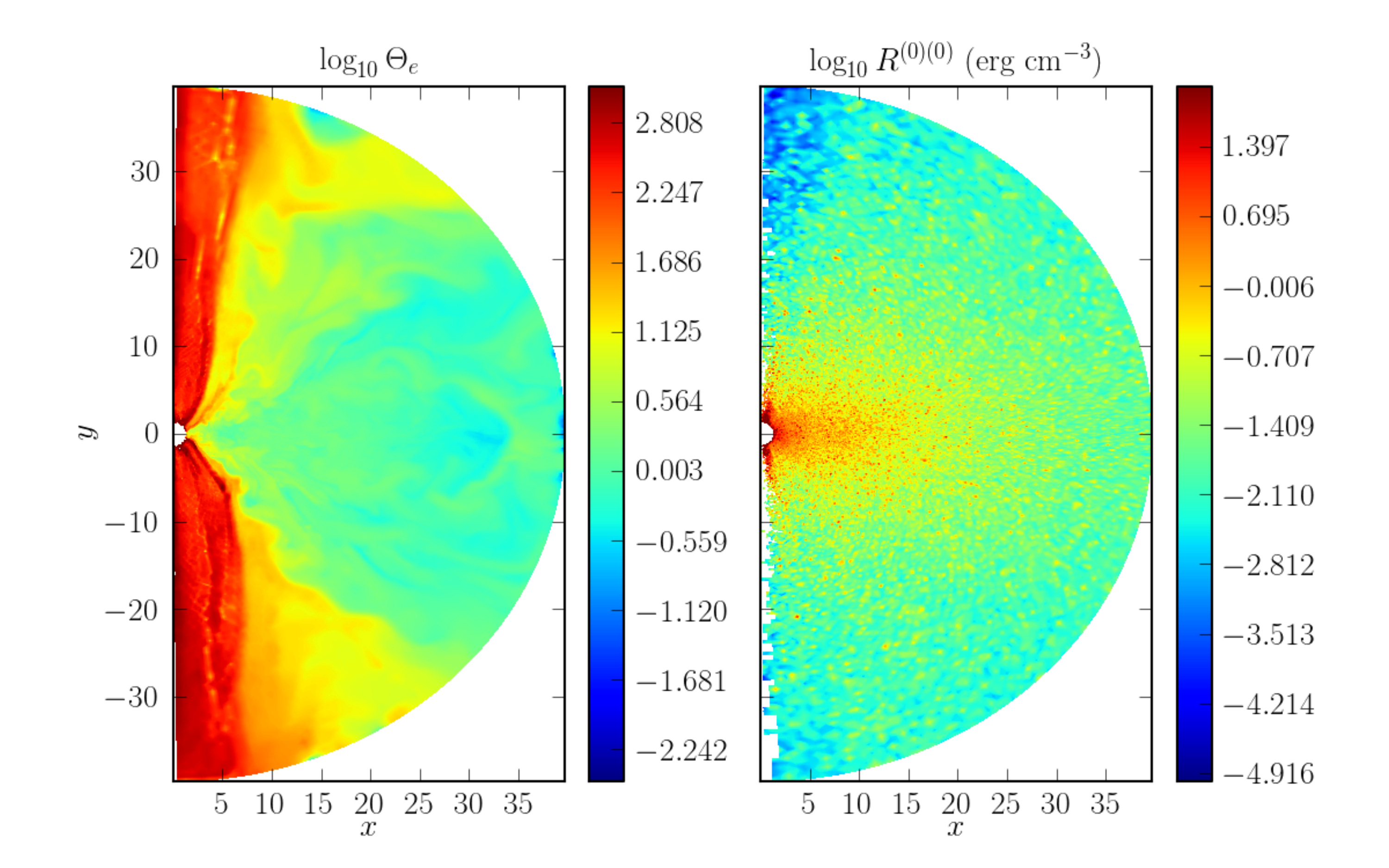}
\label{fig:torussnapshot}
\end{figure}

\begin{figure}
  \caption{Comparison of gas density $\rho$ between GRMHD and \bhl~torus calculations. Note especially that for the \bhl~result, the disk is relatively thin and the funnel is poorly developed.}
  \centering
    \plotone{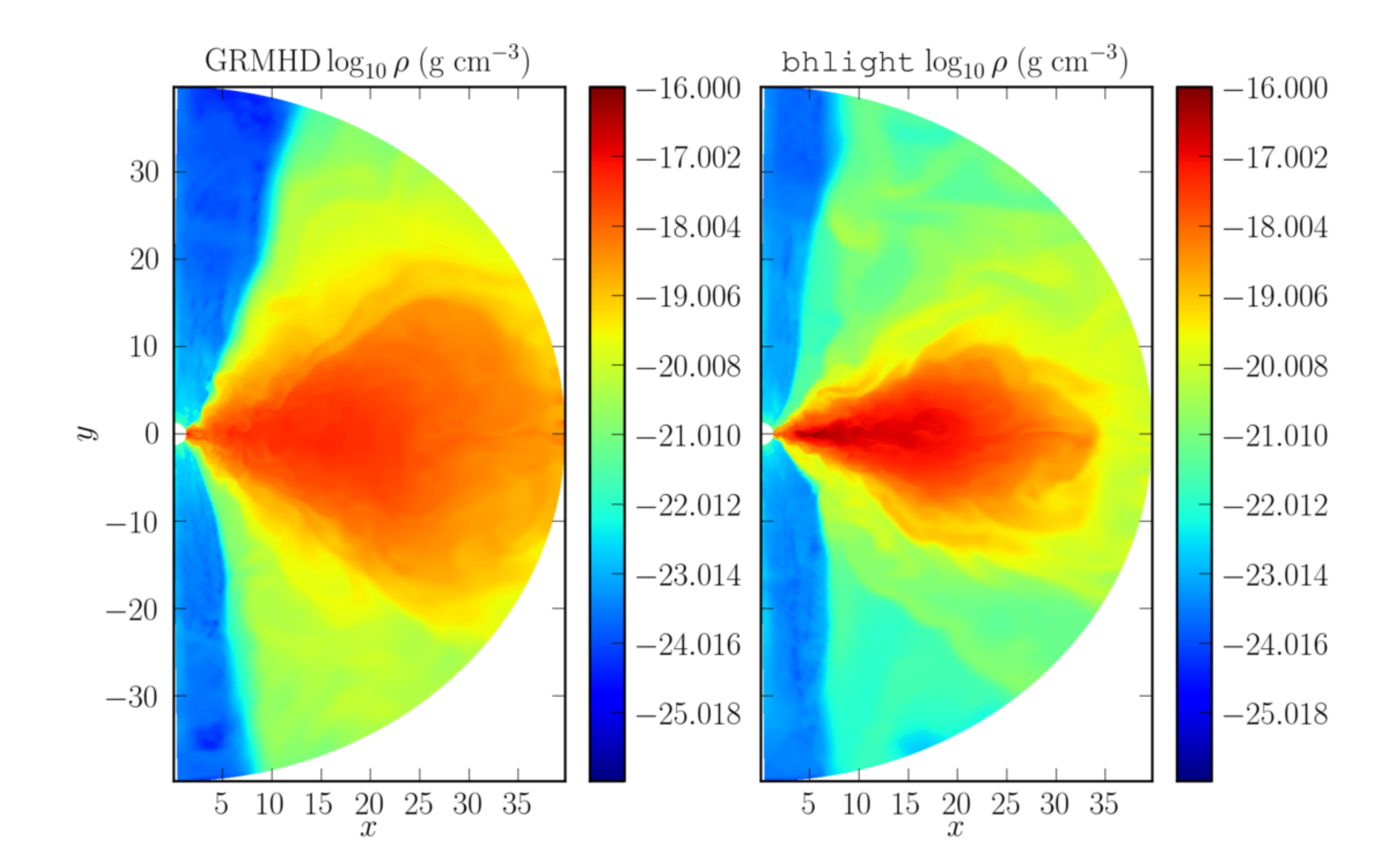}
\label{fig:torusdensitycompare}
\end{figure}

\begin{figure}
  \caption{Shell-averaged density-weighted temperature for the torus problem in \bhl~and ideal GRMHD at $t=2000M$.}
  \centering
    \plotone{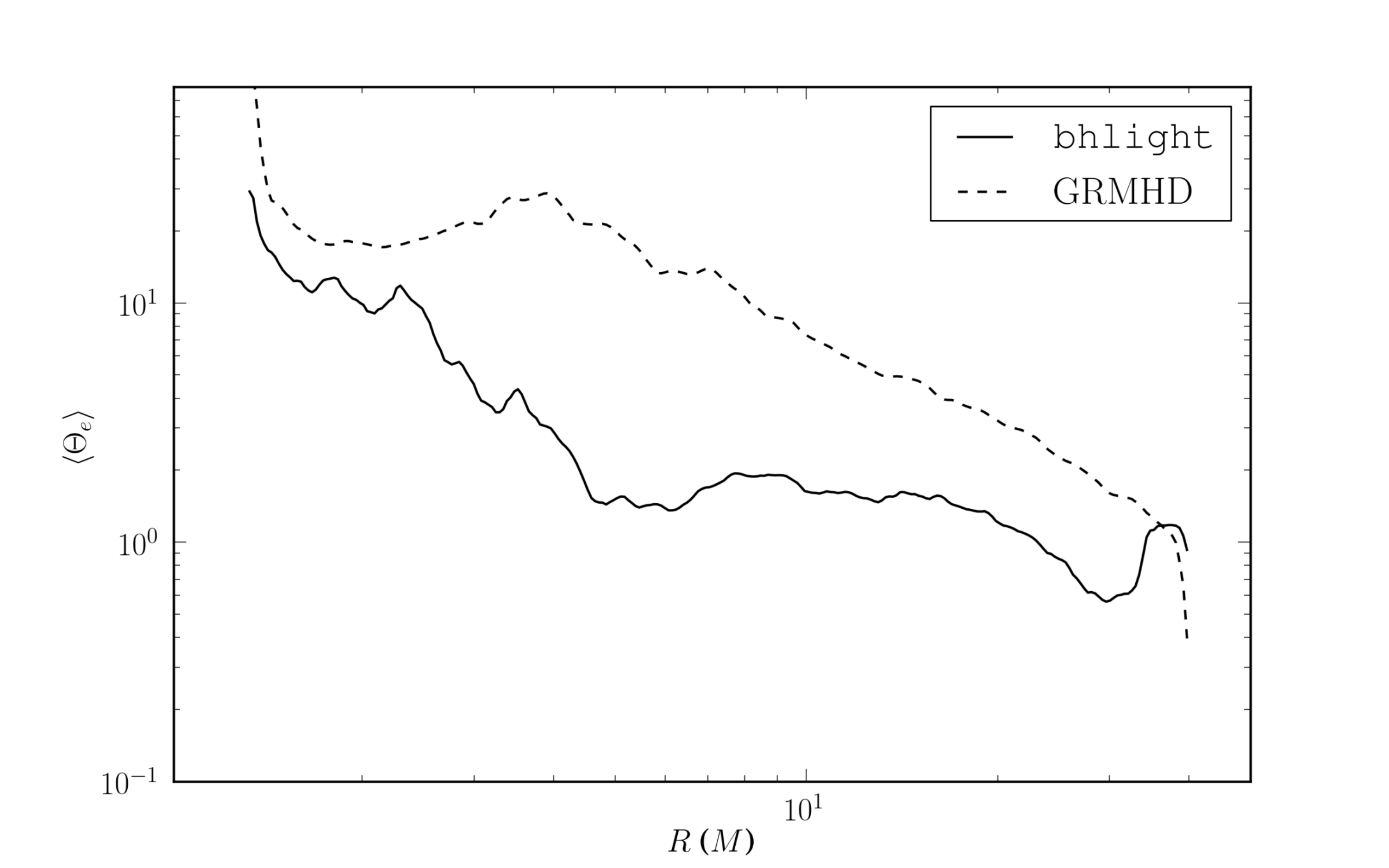}
\label{fig:sadwtorustemp}
\end{figure}

We find qualitative differences between the radiative torus simulated in \bhl~and the same model evolved with ideal GRMHD due mainly to Compton cooling in the hot, dense regions of the flow. Figure \ref{fig:torussnapshot} shows fluid temperature and comoving radiation energy density at $t = 2000M$. Figure \ref{fig:torusdensitycompare} shows density contours at $t = 2000M$ for the two models.  Figure \ref{fig:sadwtorustemp} shows shell-averaged density-weighted temperature,
\begin{equation}
\left< \Theta_e \right> \equiv \frac{\int \gdet \d x^2 \d x^3 \rho \Theta_e}{\int \gdet \d x^2 \d x^3 \rho},
\end{equation}
as a function of radius for both \bhl~and ideal GRMHD calculations; evidently the plasma is significantly cooler in the model with radiation interactions. Notice, however, that our single temperature model may be having an unrealistically strong dynamical effect; at these accretion rates the ions are likely imperfectly coupled to the electrons.

\begin{figure}
  \caption{Torus luminosity, instantaneous efficiency $\eta$, and mass accretion rate as a function of time. The dashed line denotes the thin disk efficiency at this spin. The grey region indicates the portion of $\eta$ prior to the onset of accretion across the horizon which we omit.}
  \centering
    \plotone{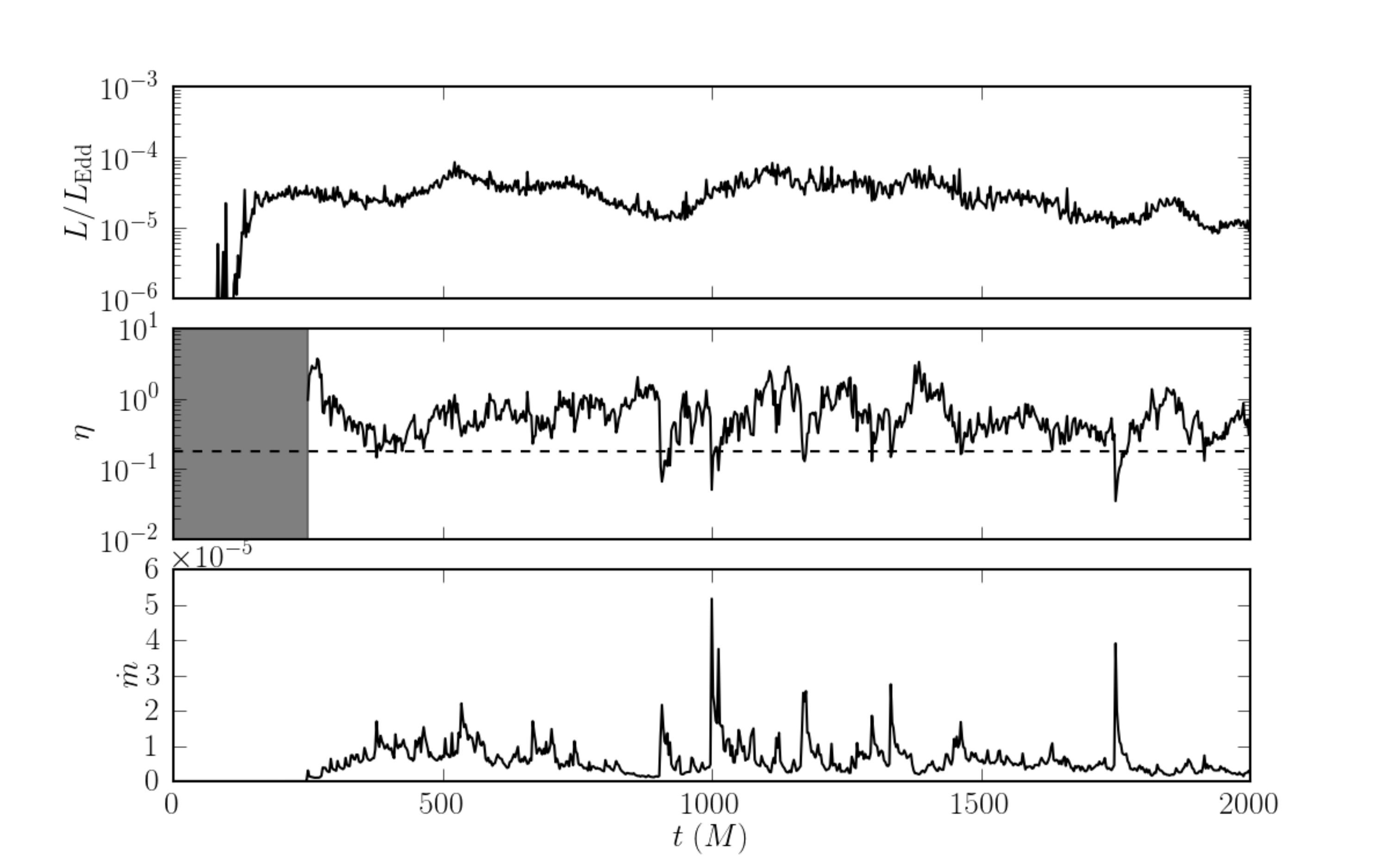}
\label{fig:torusluminosityandeffandmdot}
\end{figure}

We evaluate the luminosity $L$, Equation \ref{eqn:luminosity}, at large radius by integrating over all zones in a spherical shell.  Figure \ref{fig:torusluminosityandeffandmdot} shows $L$ and the radiative efficiency $\eta = L/\dot{M}c^2$. The mean $\eta$ we find between $t = 1750M$ and $t = 2000M$, $ \langle \eta \rangle = 0.51$, where
\begin{equation}
\langle \eta \rangle \equiv \frac{\int \d t \, L}{\int \d t \, \dot{M}c^2},
\end{equation}
 is high compared to the thin disk value, $\approx 0.18$ (for this $a_*$). Most of this energy is extracted by Compton scattering at $r \sim 10 - 15 GM/c^2$. This high efficiency is a consequence of the flow having not yet reached steady state. 

We plan to explore radiative flows at intermediate accretion rates with more sophisticated treatments of the electron physics in future work.  

\section{Conclusion}

We have introduced \bhl~which, in coupling a second order Godunov scheme to a frequency-dependent Monte Carlo radiative transfer scheme, provides a full solution to the equations of GRRMHD. \bhl~displays convergence on a number of test problems, and we have demonstrated evolution of our target application: relativistic accretion flows at moderate accretion rate. \bhl~enables more nearly {\it ab initio} study of flows in this regime, which have historically proven resistant to other methods of inquiry.

Numerical schemes have limitations. Apart from failure modes related to large optical depths and large radiation pressures mentioned previously (which prohibit near-Eddington studies as of this writing), Monte Carlo techniques are simply expensive, particularly when geodesics are nontrivial. The axisymmetric torus problem we reported was performed on one 8-core node (using {\tt mpirun} to alternate between fluid and radiation MPI sectors) for 146 hours, achieving approximately $9.5\times10^4$ superphoton updates (interactions and geodesic steps) per core-second. In comparison, the pure GRMHD torus run required only 69 core-hours, about 17 times less expensive even at this low superphoton resolution. The true minimum relative cost of \bhl~over \harm, however, will depend on the required resolution in the specific intensity for the particular problem at hand.

\acknowledgments
This work was supported by NSF grant AST 13-33612 and NASA grant NNX10AD03G, by a NASA GSRP fellowship to JCD, an Illinois Distinguished Fellowship to BRR, a Metropolis Fellowship to JCD, and a Romano Professorial Scholarship to CFG.  We thank B. Farris, C. Roedig, and particularly M. Chandra and S. Shapiro for discussions, as well as J. Stone, E. Quataert, and all the members of the horizon collaboration ({\tt horizon.astro.illinois.edu}). We also thank the anonymous referee for a very useful report. A portion of the analytic results presented here were obtained using the SageMath software package running on SageMathCloud (https://cloud.sagemath.com). A portion of the numerical results presented here were obtained on Princeton's {\tt tiger} cluster. This research is part of the Blue Waters sustained-petascale computing project, which is supported by the National Science Foundation (awards OCI-0725070 and ACI-1238993) and the state of Illinois. Blue Waters is a joint effort of the University of Illinois at Urbana-Champaign and its National Center for Supercomputing Applications.

\appendix

\section{Radiation Boundary Conditions}
In the general coordinate frame, certain boundary conditions on the superphotons required for test problems in \bhl~are more complex than in the nonrelativistic case. Here we review such boundaries as used.

\subsection{Reflecting Boundary Conditions}

The static atmosphere test in Section \ref{sec:atmosphereproblem} uses reflecting boundaries that are aligned with the coordinates.  How should one apply the reflecting boundary conditions to the radiation field?  This is not trivial in Kerr-Schild coordinates where the naive approach of simply changing the sign of the radial component of the wavevector is wrong, because the radial component of the shift does not vanish.  

When a photon crosses the boundary, we build an orthonormal tetrad with time component $e^{\mu}_{(t)} = (u^t, 0, 0, 0)$ (i.e. the tetrad is stationary in the coordinate, and hence the boundary, frame) and one spatial component that is normal to the boundary.  We transform the wavevector to the tetrad frame, reverse the normal component of the wavevector, and transform back to the coordinate frame.  

\subsection{Equilibrium Boundary Conditions}

For problems with fluid inflow across the boundary (e.g. the relativistic shocks in Section \ref{sec:radiativeshocks}) the fluid advects a thermal radiation field with it across the boundary.  How should one sample the incoming photons on the boundary?  The problem is that one is sampling a {\em flux} rather than the distribution function itself.

We have found that the simplest procedure is to sample the distribution function and multiply the weights in the sample by $\cos\theta$, where $\theta$ is the angle between the wavevector and the normal to the boundary. 

\section{The Radiating Atmosphere Under Full Transfer}
\label{apB}

We revisit the analytic solution to the radiating atmosphere problem described in Section \ref{sec:atmosphereproblem}, in which fluid and radiation confined in the Schwarzschild spacetime between two reflective spherical shells maintain a static atmosphere, without resorting to any closure for describing the radiation.

Consider the relativistic transfer equation in invariant form (ignoring scattering),
\begin{equation}
\frac{\mathrm{D}}{\mathrm{d}\lambda}\left( \frac{I_{\nu}}{\nu^3}\right) = \left( \frac{j_{\nu}}{\nu^2}\right) - \left(\nu \alpha_{\nu} \right) \left( \frac{I_{\nu}}{\nu^3}\right),
\end{equation}
which may be rewritten in terms of the invariant differential optical depth $\mathrm{d} \tau \equiv \nu \alpha_{\nu} \mathrm{d} \lambda$ (and noting that the Planck function $B_{\nu} = j_{\nu}/\alpha_{\nu}$, i.e. Kirchoff's law applies) to give
\begin{equation}
\frac{\mathrm{d}}{\mathrm{d}\tau}\left( \frac{I_{\nu}}{\nu^3}\right) = \left( \frac{B_{\nu}}{\nu^3}\right) -  \left( \frac{I_{\nu}}{\nu^3}\right),
\label{eqn:tautransfer}
\end{equation}
where all quantities in brackets are again invariant.

We adopt the ansatz for the fluid temperature distribution $\sqrt{-g_{00}}T(r) =T_{\infty}$ from Section \ref{sec:atmosphereproblem}. Consider also the frequency $\omega = -k_{\mu}u^{\mu}$ of a superphoton at radius $r$ in the Schwarzschild spacetime. The four-velocity of a local frame not moving with respect to the coordinate system is $u^{\mu} = (1/\sqrt{-g_{00}},0,0,0)$: then, $\omega = -k_{0}/\sqrt{-g_{00}}$. The invariant Planck function $B_{\nu}/ \nu^3 =(h^4 /c^2)f(h\nu/k_B T)$. However, $h\nu/k_B T = -h k_0/2 \pi k_B T_{\infty}$ is constant along geodesics;  $B_{\nu}/ \nu^3$ is therefore also constant along every ray. 

For reflecting boundary conditions, rays of the intensity $I_{\nu}$ do not terminate; they instead repeatedly reflect off the boundaries all the way back to $\mathrm{d}\tau \rightarrow \infty$. Because $B_{\nu}/ \nu^3$ is constant along every ray, for the stationary system we have $I_{\nu}/ \nu^3 = B_{\nu}/ \nu^3$ everywhere. Our assumed temperature distribution is therefore consistent with the full transfer equation, and the solution presented in Section \ref{sec:atmosphereproblem} is exact for all optical depths. 

\section{Linear Modes in Relativistic Radiation Magnetohydrodynamics}
\label{linearmodes}

We present the linearized equations of radiation magnetohydrodynamics in flat spacetime, assuming the Eddington closure \citep{farris2008} with a gray absorption opacity $\kappa$ and setting $k_B = c=1$. The governing equations are then given in terms of divergences of the matter four-current and the MHD and radiation stress-energy tensors, along with the magnetic induction equation. In plane-parallel symmetry, we search for wave solutions of the form ${\bf P} = (\rho, u, u^1, u^2, B^1, B^2, E, F^1, F^2) = (\rho_0 + \delta \rho, u_0 + \delta u, \delta u^1, \delta u^2, B_0, B_0 + \delta B^2, E_0 + \delta E, \delta F^1, \delta F^2)$ (i.e.~we confine variation to a plane), where $\delta \propto \exp(\omega t - i k x)$ is a small perturbation, and $E_0 = a_R ((\gamma-1)u_0/\rho_0)^4$ to enforce radiative equilibrium of the background state. We write the linearized systems in the form $\omega \delta {\bf P} = {\bf A} \delta {\bf P}$; the dispersion relation is then $\mathrm{det}({\bf A} - \mathbb{I}\omega)=0$. We find that the matrix ${\bf A}$ is

\begin{equation*}
\left(
\begin{array}{cccccccc}
%[0, N]
0&0&-i  k \rho_{0}&0&0&0&0&0 \\
%[1, N]
4 \kappa E_0&\frac{-4 \kappa E_0 \rho_0}{u_0}&-i k\gamma  u_0&0&0&\kappa \rho_0&0&0 \\
%[2, N]
0&\frac{-ik\left( \gamma-1\right) C_2}{C_{1}}&0&0&\frac{-i B_{0} k}{C_2 + B_0^2}&0&\frac{\kappa \rho_0 C_2}{C_{1}}&\frac{B_{0}^{2} \kappa \rho_{0}}{C_{1}} \\
%[3, N]
0&\frac{-ik{\left( \gamma - 1\right)}B_{0}^{2} }{C_{1}}&0&0&\frac{i B_{0} k}{C_2 + B_0^2}&0&\frac{B_{0}^{2} \kappa \rho_{0}}{C_{1}}&\frac{\kappa \rho_0 C_2}{C_{1}} \\
%[4, N]
0&0&-i k B_{0} &i k B_{0} &0&0&0&0 \\
%[5, N]
-4 \kappa E_{0}&\frac{4 E_{0} \kappa \rho_{0}}{u_{0}}&-\frac{4}{3} i k E_{0} &0&0&-\kappa \rho_{0}&-i k&0 \\
%[6, N]
0&\frac{4ikC_2 E_{0}{\left(\gamma - 1\right)}}{3 C_{1}}&0&0&\frac{4 i k B_{0} E_{0}}{3  {\left(C_2 + B_0^2\right)}}&-\frac{1}{3} i k&\frac{-C_{3} \kappa \rho_{0}}{3 C_{1}}& \frac{-4 B_{0}^{2} E_{0} \kappa \rho_{0}}{3 C_{1}}\\
%[7, N]
0&\frac{4 i k B_{0}^{2} E_{0} {\left(\gamma - 1\right)}}{3  C_{1}}&0&0&\frac{-4 i k B_{0} E_{0}}{3 {\left(C_2 + B_0^2\right)}}&0&\frac{-4 \kappa \rho_0 B_{0}^{2} E_{0}}{3  C_{1}}& \frac{-C_{3} \kappa \rho_{0}}{3  C_{1}}\\
\end{array}
\right),
\end{equation*}
where 
\begin{align*}
C_1 &= {\gamma^{2} u_{0}^{2} + 2 {\left(\gamma u_{0} + \rho_{0}\right)} B_{0}^{2} + 2 \gamma \rho_{0} u_{0} + \rho_{0}^{2}}, \\
C_2 &= {B_{0}^{2} + \gamma u_{0} + \rho_{0}}, \\
%C_3 &= 3 C_1 + 4B_0^2E_0 + 4 \Gamma E_0u_0 + 4 E_0 \rho_0.
C_3 &= 3 C_1 + 4 E_0 C_2.
\end{align*}

\bibliographystyle{apj}
\bibliography{local}

\end{document}